\documentclass[journal,preprint]{vgtc}                

\usepackage{mathptmx}
\usepackage{graphicx}
\usepackage{times}
\usepackage{amsmath,amssymb}
\usepackage{algorithm,algpseudocode}
\usepackage{multirow}
\usepackage{caption}
\usepackage[mathcal]{euscript}
\usepackage{stmaryrd}
\usepackage{tikz}
\usetikzlibrary{shapes.misc}
\usetikzlibrary{decorations.pathreplacing}
\usepackage{mathtools}
\usepackage{empheq}

\usepackage{color,soul}
\definecolor{clr1}{RGB}{103,0,31}
\definecolor{clr2}{RGB}{178,24,43}
\definecolor{clr3}{RGB}{214,96,77}
\definecolor{clr4}{RGB}{244,165,130}
\definecolor{clr5}{RGB}{253,219,199}
\definecolor{clr6}{RGB}{247,247,247}
\definecolor{clr7}{RGB}{209,229,240}
\definecolor{clr8}{RGB}{146,197,222}
\definecolor{clr9}{RGB}{67,147,195}
\definecolor{clr10}{RGB}{33,102,172}
\definecolor{clr11}{RGB}{5,48,97}
\definecolor{white}{RGB}{255,255,255}

\renewcommand{\hl}[1]{#1}
\newcommand{\hlm}{\mbox}
\newcommand{\note}{}

\newcommand{\IGNORE}[1]{}

\newcommand{\N}{\ensuremath{\mathbb{N}}}
\newcommand{\R}{\ensuremath{\mathbb{R}}}
\newcommand{\Seg}{\ensuremath{\mathcal{S}}}
\newcommand{\His}{\ensuremath{\mathcal{H}}}
\newcommand{\Dom}{\ensuremath{\mathcal{D}}}
\newcommand{\Mix}{\ensuremath{\mathcal{M}}}

\newcommand{\G}{\ensuremath{\mathcal{G}}}
\newcommand{\vc}[1]{\ensuremath{\mathbf{#1}}}
\newcommand{\unit}[1]{\ensuremath{\widehat{\vc{e}}_{#1}}}
\newcommand{\comp}[2]{\ensuremath{\left[\,\vc{#1}\,\right]_{#2}}}
\newcommand{\maxc}[1]{\ensuremath{\widehat{k}\left(\vc{#1}\right)}}

\DeclareMathOperator*{\argmax}{arg\,max}

\usepackage[bookmarks,backref=true,linkcolor=black]{hyperref} 
\hypersetup{
  pdfauthor = {},
  pdftitle = {},
  pdfsubject = {},
  pdfkeywords = {},
  colorlinks=true,
  linkcolor= black,
  citecolor= black,
  pageanchor=true,
  urlcolor = black,
  plainpages = false,
  linktocpage
}

\onlineid{1038}

\vgtccategory{Research}

\vgtcinsertpkg

\preprinttext{To appear in IEEE Transactions on Visualization and Computer Graphics (IEEE Vis 2020).}

\title{The Mixture Graph---A Data Structure for Compressing, Rendering, and Querying Segmentation Histograms}

\author{Khaled Al-Thelaya~~~~~~~Marco Agus~~~~~~~Jens Schneider (\textit{IEEE Member}) }
\authorfooter{
\item
 All authors are with the College of Science and Engineering (CSE), Hamad Bin Khalifa University (HBKU), Education City, Doha, Qatar.
\item
 Jens Schneider is the corresponding author and can be reached at jeschneider@hbku.edu.qa.
}

\shortauthortitle{Al-Thelaya \MakeLowercase{\textit{et al.}}: The Mixture Graph}

\abstract{In this paper, we present a novel data structure, called the \textit{Mixture Graph}. This data structure allows us to compress, render, and query segmentation histograms. Such histograms arise when building a mipmap of a volume containing segmentation IDs. Each voxel in the histogram mipmap contains a convex combination (\textsl{mixture}) of segmentation IDs. Each mixture represents the distribution of IDs in the respective voxel's children. Our method factorizes these mixtures into a series of linear interpolations between exactly two segmentation IDs. The result is represented as a directed acyclic graph (DAG) whose nodes are topologically ordered. Pruning replicate nodes in the tree followed by compression allows us to store the resulting data structure efficiently. During rendering, transfer functions are propagated from sources (leafs) through the DAG to allow for efficient, pre-filtered rendering \hl{at interactive frame rates}. Assembly of histogram contributions across the footprint of a given volume allows us to efficiently query partial histograms\hl{, achieving up to $178\times$ speed-up over na\"{i}ve parallelized range queries.} Additionally, we apply the Mixture Graph to compute correctly pre-filtered volume lighting \hl{and to interactively explore segments based on shape, geometry, and orientation using multi-dimensional transfer functions.}
} 


\teaser{
\centering

	\includegraphics[height=3.5cm]{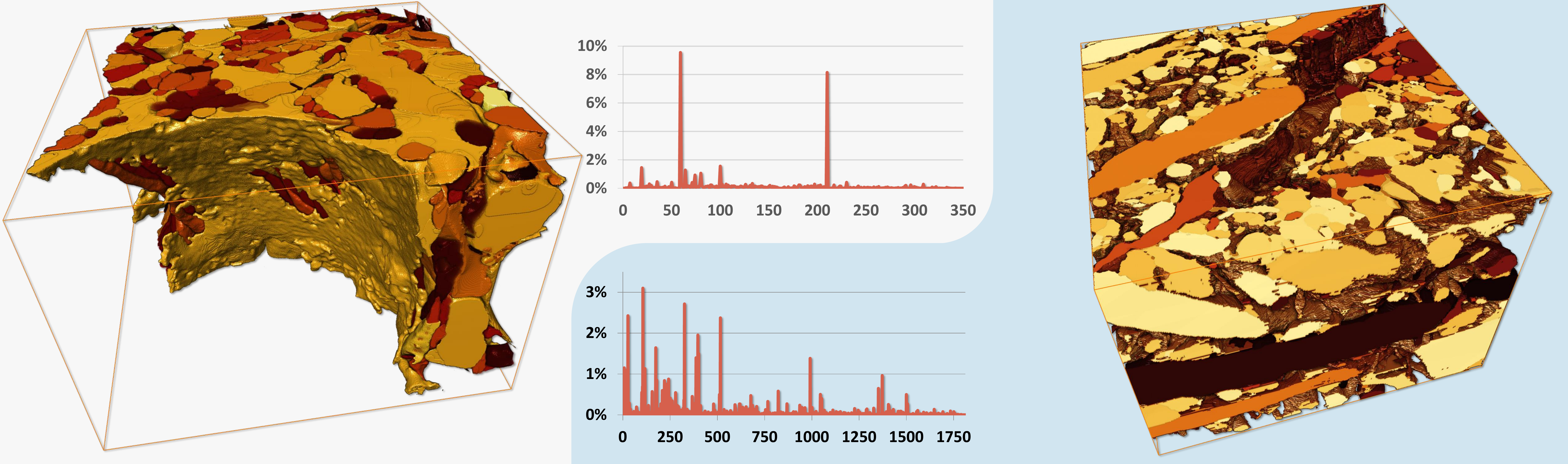}
	\includegraphics[height=3.5cm]{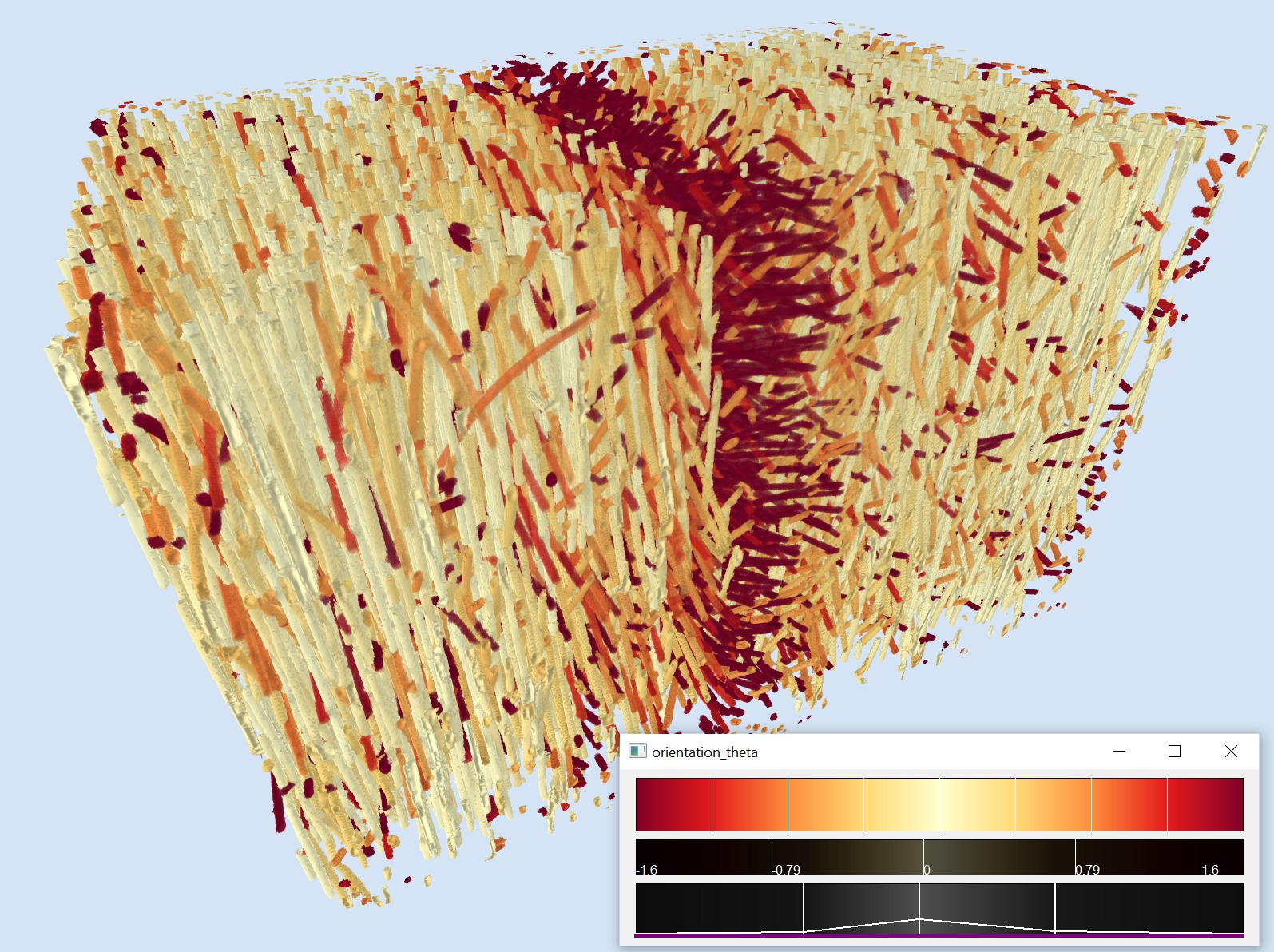}
	\vspace{-2mm}
	\caption{\label{fig:teaser}	\hl{The \textit{Mixture Graph} allows us to interactively render and query segmented volumes while maintaining pre-filtered anti-aliasing across multiple scales. Segmented}  volumes store nominal data and therefore pose challenges if traditional rendering is attempted. \textbf{Left to right:}
	Hippocampus ($1178\times1125\times789$, 353 segments), Neocortex ($2048\times 2048\times 300$, 1,182 segments), fiber-reinforced Polymer ($614\times 961 \times 600$, 15,917 segments). 
	}
}

\begin{document}
\shortauthortitle{{Al-Thelaya, Agus, Schneider}: The Mixture Graph}

\firstsection{Introduction}
\maketitle
Segmentations of volume data have traditionally been used in a medical context to separate different semantic objects in the data. 
Typically, one or more segment IDs are computed per voxel that represent the probability of the voxel belonging to the specific segment.
Such segmentations have then been used either to modulate the transfer function or to extract 3D representations for each semantic object. 
Since then, segmented volumes have also become popular in other disciplines, such as engineering, e.g., in the form of topological segmentation of flow fields. In \hl{all} these \hl{examples}, however, segmentations have been primarily treated as annotations to the original data.

This is changing due to the extensive use of segmentations in fields such as neuroscience, connectomics (see also Fig.~\ref{fig:teaser}, left \& middle), computational neurology and material science (see also Fig.~\ref{fig:teaser}, right). 
\hl{Here}, raw data is typically imaged using \hl{an} electron microscope (EM). This data, however, primarily serves as a means to compute the segmentation. Once the raw volume has been successfully partitioned into semantic objects, the raw data is used less and less. Reasons for this procedure include that the raw EM data is noisy and often lacks a direct optical interpretation. Some of the data in the OpenConnectome Project~\cite{OpenConnectome}, for instance, is imaged using a high-resolution scanning (non-transmissive) EM~\cite{Kasthuri15}. Such data does represent \hl{neither opacity nor color}. \hl{Therefore, this traditional input for volume rendering can be assigned using transfer functions~{\cite{Ljung16}} prior to rendering.} Alternatively, 3D surface representations of semantic objects may be used. Still, volume rendering of the raw data~\cite{Jeong10b} is highly regarded as a valuable tool in order to obtain, author, and annotate~\cite{Beyer13,Beyer13b}, or validate~\cite{Awami15} the resulting segmentation. However, as segmentation volumes are now a \textit{primary} modality rather than a derived quantity, interactive, direct rendering of segmentations gains importance~\hl{{\cite{Beyer:2019:CESSV}}}.

Such segmentation volumes store \textit{nominal} data, posing challenges to traditional rendering attempts \hl{such as:} data cannot be interpolated \hl{or pre-filtered} before \hl{assigning optical properties to voxels. This implies that, traditionally, mipmaps~{\cite{Williams83}} need to be recomputed from scratch whenever a change in the transfer function alters these voxel properties.}
Adding to the challenges, volumes generated in connectomics are among the largest volumetric data sets currently available, \hl{making pre-filtered anti-aliasing highly desireable. Moreover, the nominal character of the data also makes reducing data size while maintaining direct renderability difficult}, ruling out virtually all existing lossy compression methods.

\noindent\textbf{Contributions.} In this paper, we address some of the most dire of the aforementioned needs. In particular, we present the Mixture Graph, a novel, \hl{graph-based data structure to represent hierarchical segmentation histograms. We show that these histograms can be understood as computational instruction for building a mipmap to support pre-filtered anti-aliasing \textit{before} the optical properties at voxels are known. Compared to na\"{i}ve use of such histograms, the Mixture Graph presents an efficient and compact alternative and allows us to propagate transfer function updates efficiently and in parallel through the graph. This is achieved by a symbolic factorization that breaks the aforementioned computational instruction into a sequence of linear interpolations. These interpolations can be compressed using well-established, lossy compression techniques. This leads to a prune-and-recycle operation on the underlying graph that exploits sparsity in the histogram's range.}
The resulting data structure naturally supports pre-filtered rendering. We also describe how the Mixture Graph can be used to store quantized normals to provide simple pre-filtered shading. Finally, we present a footprint assembly algorithm to efficiently compute partial segment histograms across a given sub-volume.

\section{Related Work}
In connectomics and material sciences, segmentation volumes have ceased to be mere annotations to the semantic objects in the data. Instead, the segmentation has become a first-class modality to be analyzed and visualized. Reasons include that connectomics data is typically imaged with an electron microscope, an imaging modality prone to noise which does not necessarily result in data lending itself to interpretation in terms of optical absorption and opacity. For instance, unlike transmission EMs, the more commonly used scanning EMs measure the backscattering of the electron beam~\cite{Schalek16}. Still, direct volume rendering is used~\cite{Hadwiger12,Jeong10b}, often in combination with other techniques, to proofread~\cite{Haehn14, Awami15} and analyze or explore~\cite{Awami14,Beyer13b,Beyer13,Jeong10} automatically~\cite{Kaynig15,Vazquez11,Roncal15} or semi-automatically~\cite{Jeong09} generated segmentations.

As more and more automated segmentation algorithms become available, large and densely segmented data sets emerge. Unlike the quantized real values stored by their EM input, these data sets store nominal integer IDs. Rendering such modalities was addressed, e.g., by two-level volume rendering~\cite{Hauser01,Hadwiger03}, which combines multiple volume rendering techniques 
to highlight the various semantic objects in combination with the original input data. Other approaches seek to reconstruct a smooth surface for purposes of rendering~\cite{Lempitsky:10}. 

While all this demonstrates the interest in segmented data, compression of \textit{segmented volume data} gained relatively little attention. The reason is that, albeit, being the most popular and successful choice in the last decades, lossy compression methods~\cite{Ihm99, Nguyen01, Schneider03} cannot be readily used to compress nominal integer data. On the other hand, lossless compression methods face challenges with respect to implementation and performance on parallel architectures~\cite{Owens12}. \hl{This is due to their inherently sequential view of the data, and, while such methods exist~}\cite{Funasaka:2010,Weissenberger:2019,Weissenberger:2018,Sitardi:2016}\hl{, they generally do offer neither the random access nor the bandwidth required to interactively render directly from the compressed representation.} Aside from raw integer volumes~\cite{OpenConnectome}, PNG-compressed RGB or RGB$\alpha$ slices storing the segment IDs in multiple 8-bit channels are still a de-facto format for exchanging columns of neuronal tissue~\cite{Kasthuri15, Cali16}. PNG's compression ratio for this type of data is generally significant (e.g., $<0.3$ bits per pixel for the Hippocampus~\cite{Cali16}, Fig.~\ref{fig:teaser} left), reflecting the sparse nature of segmentation data. However, such a compression does not address the generation of hierarchies from the data. \hl{Such hierarchies (mipmaps) are crucial in providing pre-filtered anti-aliasing for high resolution volumes}. Furthermore, PNG stacks need to be decompressed prior to rendering, \hl{e.g., to synthesize views oblique to the slices. Unlike traditional medical segmentations, the number of segment IDs in connectomics and material sciences that individual binary segmentations have little to no practical relevance.}

Traditional compression methods for \textit{scalar volume data} focus on reducing the size of opacity values. Wavelets~\cite{Daubechies1998, Cohen92}, while resulting in good compression rates~\cite{Muraki93, Ihm99, Nguyen01, Bajaj01} are usually not well suited for decoding on the GPU. The reason is that they derive their efficiency not only from the actual wavelet transform, but from the coding back-end of the transform coefficients. Arithmetic codes~\cite{arith} or variations of embedded zero tree codes (e.g., SPIHT~\cite{Said96}) are traditionally used. These do not trivially support random parallel access. GPU-based wavelets are employed in the field of terrain rendering~\cite{Treib12} and octrees~\cite{octree} are used to compress volume data. All these methods, however, do not support lossless encoding of nominal data such as the segmentation volumes we are concerned with. While lossless coding may also be driven by a wavelet transform, such as the fully invertible LeGall integer basis~\cite{Cohen92}. To the best of our knowledge, however, such lossless transforms have not yet been applied to volume data.

Vector quantization~\cite{Gersho} has been applied successfully to the lossy compression of volume data~\cite{Ning92, Schneider03, Fout07}. Vector quantization is similar to our method in that a palette or codebook is learned from the data. Each entry in the codebook stores a vector, whereas each vector in the input data is replaced by an index into the codebook. In this work, we utilize vector quantization to derive a nominal volume from a normal map to apply the Mixture Graph for pre-filtered shading. The related field of sparse coding and sparse dictionary learning~\cite{Rubinstein10} can be seen as a generalization of vector quantization: instead of referencing the codebook with a single index, a sparse weight vector is stored. Decoding consists of computing a linear combination of codebook entries.
\hl{More recently, Wang et al.~}\cite{Wang:2017, Wang:2018} \hl{propose to  apply sparse 3D Gaussian Mixture Models to handle massive scalar simulation volumes.}

In this work, we consider the efficient, hierarchical storage of segmented volume data. Our method considers a mipmap of attributes, such as segment color, that would arise under application of a transfer function to the input segmentation. However, unlike a traditional mipmap that is built \textsl{after} a transfer function is applied, our data structure stores the \textsl{computations} necessary to arrive at a mipmap. These computations are factorized into simple linear interpolations that can be compressed efficiently and by lossy methods. Our data representation is essentially a paletted texture, in which the palette can be updated efficiently and in parallel. Unlike traditional paletted textures, however, our palette can store mixtures of multiple colors, somewhat similar to two-colored pixels~\cite{Pavic10}. Our method is \hl{most closely} related to double sparsity and sparse coding~\cite{Rubinstein10}, although the methodology differs substantially: We perform a greedy factorization of the \textsl{computations} necessary to compute a palette, whereas sparse coding is usually formulated as an (orthogonal) matching pursuit \hl{ {\cite{Mallat:1993:MP}} } optimization problem to represent data in a potentially overcomplete basis.

\section{Algorithmic Overview}
Our method first computes a normalized histogram mipmap (Sec.~\ref{sec:histo}), in which each voxel stores a ``mixture'' of segment IDs. 
Mixtures are convex combinations of segment IDs reflecting the relative number of occurrences of each ID within each voxel of the mipmap. Clearly, such a histogram may have significant storage requirements and typically results in heavily unbalanced workloads during rendering.

In a second step, we factorize the histogram mipmap into a set of linear interpolations (Sec.~\ref{sec:factor}). Made possible by embedding mixtures in $\R^\infty$, this step results in a directed, acyclic graph (DAG) representing both the histogram \textsl{and} the computations necessary to reconstruct or render the original segmented volume at different scales (Sec.~\ref{sec:reconstruct}).

We use a scalar quantization step to prune redundant nodes in the DAG. 
The result is then compressed at a fixed bitrate to facilitate fast, random access to the original histogram (Sec.~\ref{sec:compression}). 

Since the quantization step is lossy, we also store quantization errors for each quantization bin to estimate the reconstruction error in later stages (Sec.~\ref{sec:error}), such as running fast, approximate queries of partial histograms across any given sub-volume. This feature allows domain scientists to quickly count IDs in a volumetric range and assess their distribution (Sec.~\ref{sec:svq}).\\

\noindent\textbf{Notation.} 
In this paper, we make heavy use of convex combinations that we call ``mixtures''. A mixture is described by a vector $\vc{m}\in\R^\infty$ with the following properties.
\begin{align}
	\label{eq:convex1}\left\|\vc{m}\right\|_1 &=1\\
	\label{eq:convex2}\comp{m}{n}\, &\geq 0\;\forall n\\
	\label{eq:finite}\left\|\vc{m}\right\|_0 & < \infty,
\end{align}
where we used the $\ell_0$ pseudo-norm to count non-zero elements in $\vc{m}$ and used $\comp{\cdot}{n}$ to denote the $n^\mathrm{th}$ element in a vector. The scalar product $\langle\vc{m},\vc{x}\rangle$
between a mixture vector $\vc{m}$ and a vector $\vc{x}\in\R^\infty$ storing data thus computes a convex combination (Eq.~(\ref{eq:convex1},\ref{eq:convex2})) of a finite number of elements (Eq.~\eqref{eq:finite}) in $\vc{x}$. We further define the set of all mixtures
\begin{align}
	\Mix &:=\left\{\;\vc{m}\in\R^\infty\; :\; \|\vc{m}\|_1 = 1\;\; \wedge\;\; \|\vc{m}\|_0< \infty\;\; \wedge\;\; \comp{m}{n}\geq 0\;\forall n\;\right\}.\nonumber
\end{align}
Finally, we generally assume that the greatest position of a non-zero element in $\vc{m}$ is known and finite, denoted\vspace{-1mm}
\begin{align}
	\exists\, \maxc{m}\in\N_0 &: \comp{m}{k} = 0\quad \forall k>\maxc{m}.\nonumber
\end{align}

\subsection{Normalized Histogram Mipmap}
\label{sec:histo}
Given a compact, discrete domain, $\Dom \subseteq \N_0^3$, and a volume of segment IDs,
$\Seg : \Dom \rightarrow \N_0$, we construct a hierarchical segmentation histogram with \hl{$l_{\max}+1$} levels, 
$\His : \Dom\times\left\{0,\ldots,l_{\max}\right\} \rightarrow \Mix$ as follows. Let $i,j,k \in \Dom$ denote a voxel position in 3D and $l\in\left\{0,\ldots,l_{\max}\right\}$ denote a hierarchy level. Let $l=0$ refer to the level with the finest resolution and let $\unit{i}$ denote the $i^\mathrm{th}$ unit vector over $\R^\infty$. We then compute
\begin{align}\vspace{-1mm}
	\label{eq:mip}
	\His(i,j,k,0) &= \unit{\Seg(i,j,k)}\quad\forall i,j,k \in \Dom\nonumber\\
	\His(i,j,k,l) &= \frac{1}{N}~\sum_{u=2i}^{2i+1}\;\;\sum_{v=2j}^{2j+1}\;\;\sum_{w=2k}^{2k+1}\;\; \His(u,v,w,l-1),
\end{align}
where $N$ is the number of voxels in the support of $\His(i,j,k,l)$ (typically 8, but potentially less than 8 at the borders). If the input data has non-power-of-two resolution, we round up the resolution of each subsequent level, adjusting summation limits and $N$ in Eq.~\eqref{eq:mip} accordingly. We continue computing additional levels in $\His$ in this way until we reach $l = l_{\max}$ with a resolution of $1^3$ voxels.

Our assumption that the input volume $\Seg$ contains exactly one segment ID per voxel merely served the exposition of this section. If voxels in $\Seg$ already contain mixtures (i.e., $\Seg:\Dom \rightarrow\Mix$), we set $\His(\cdot,0)=\Seg$ and proceed as described above. In this paper, we will only discuss the traditional \textsl{average-of-eight} mipmap filter~\cite{Williams83}, but other low-pass filters can be used in the construction of $\His$, as long as the low-pass filter can be normalized to a mixture itself.

\subsection{Factorization}
\label{sec:factor}
High-resolution levels of $\His$ are very sparse for most real-world segmented volumes. This is particularly true for volumes in which only one segment ID is provided per voxel. In contrast, $\His(\cdot,l_{\max})$ is the dense, normalized histogram of all segment IDs in the volume. This poses challenges for processing and rendering such histograms, since the workload per voxel is highly inhomogeneous: In order to render a segmented volume with 1,024 IDs using an RGB$\alpha$ transfer function, only one fetch is sufficient for each voxel in $l=0$, whereas higher levels require up to 1,024 fetches to compute the color of a single voxel. To balance this workload, we propose to factorize each mixture into a set of ``simpler'' mixtures. In this paper, we consider mixtures of the form $\vc{\lambda}\in\Mix\,:\,\|\vc{\lambda}\|_0\leq 2$, that is, we restrict the factorization to either linear interpolation between two elements or identity of one element.

Such a factorization is always possible in the $\R^\infty$ embedding.
We start by populating a mixture list $\Lambda$ with $N$ trivial mixtures, one for each input segment: $\Lambda=\left[\;\unit{1},\ldots,\unit{N}\;\right]$. We then examine one of the remaining mixtures $\vc{m}$ ``over $\Lambda$'' (that is, $\comp{m}{n}\neq0\Leftrightarrow n\in\left\{1,\ldots,N\right\}$) with $\|\vc{m}\|_0>2$. We pick two non-zero positions $i,j$ (i.e., $\comp{m}{i}\neq 0$ and $\comp{m}{j}\neq 0$). After that, we compute a new mixture\vspace{-1mm}
\begin{align}
	\vc{\lambda}_{N+1}&:= \frac{\comp{m}{i}\unit{i} + \comp{m}{j}\unit{j}}{\comp{m}{i} + \comp{m}{j}},
\end{align}
which is appended to the mixture list\vspace{-1mm}
\begin{align}
	\Lambda &\mapsfrom \Lambda \oplus \vc{\lambda}_{N+1}.
\end{align}
\begin{figure}[t]
\centering
	\fbox{
		\begin{minipage}{0.75\columnwidth}
		\footnotesize
		\renewcommand{\arraystretch}{1.2}
		\centering
		\begin{tabular}{ll}
			Input:     & $\vc{m}=\left(\;0.1,0.2,0.3,0.4,\ldots\;\right)$\\
			Mixture List: & $\Lambda=\left[\;\unit{1},\unit{2},\unit{3},\unit{4}\;\right]$\vspace{0.5mm}\\
			\hline\\[-2.5ex]
			Pick ${\color{clr1}\mathbf{1}},{\color{clr3}\mathbf{2}}$:  & $\vc{m}=\left(\;{\color{clr1}\mathbf{0.1}},{\color{clr3}\mathbf{0.2}},0.3,0.4,\ldots\;\right)$\\
			Insert new ${\color{clr10}\mathbf{\lambda}}$: &
			{\tabcolsep0pt
				\begin{tabular}{l}
					$\Lambda = \left[\;\unit{1},\unit{2},\unit{3},\unit{4},{\color{clr10}\mathbf{\vc{\lambda}_5}}\;\right]$,\\
					${\color{clr10}\mathbf{\vc{\lambda}_5}}=\frac{1}{3}\left(\;{\color{clr1}\mathbf{1}},{\color{clr3}\mathbf{2}},\ldots\;\right)$					
				\end{tabular}
			}\\
		Update: & $\vc{m}=\left(\;{\color{clr10}{\mathbf{0}}},{\color{clr10}{\mathbf{0}}},0.3,0.4,{\color{clr10}{\mathbf{0.3}}},\ldots\;\right)$\vspace{0.5mm}\\
		\hline\\[-2.5ex]
			Pick ${\color{clr1}\mathbf{3}},{\color{clr3}\mathbf{5}}$: & $\vc{m}=\left(\;0,0,{\color{clr1}\mathbf{0.3}},0.4,{\color{clr3}\mathbf{0.3}},\ldots\;\right)$\\
			Insert new ${\color{clr10}\mathbf{\lambda}}$: &
			{\tabcolsep0pt
				\begin{tabular}{l}
					$\Lambda = \left[\;\unit{1},\unit{2},\unit{3},\unit{4},\vc{\lambda}_5,{\color{clr10}\mathbf{\vc{\lambda}_6}}\;\right]$,\\
			${\color{clr10}\mathbf{\vc{\lambda}_6}}=\frac{1}{2}\left(\;0,0,{\color{clr1}\mathbf{1}},0,{\color{clr3}\mathbf{1}},\ldots\;\right)$\\					
				\end{tabular}
			}\\
		Update: & $\vc{m} = \left(\;0,0,{\color{clr10}\mathbf{0}},0.4,{\color{clr10}\mathbf{0}},{\color{clr10}\mathbf{0.6}},\ldots\;\right)$\vspace{0.5mm}\\
		\hline\\[-2.5ex]
		
		Pick ${\color{clr1}\mathbf{4}},{\color{clr3}\mathbf{6}}$: & $\vc{m}=\left(\;0,0,0,{\color{clr1}\mathbf{0.4}},0,{\color{clr3}\mathbf{0.6}},\ldots\;\right)$\\
		Insert new ${\color{clr10}\mathbf{\lambda}}$: &
		{\tabcolsep0pt
			\begin{tabular}{l}
				$\Lambda = \left[\;\unit{1},\unit{2},\unit{3},\unit{4},\vc{\lambda}_5,\vc{\lambda}_6,{\color{clr10}\mathbf{\vc{\lambda}_7}}\;\right]$,\\
				${\color{clr10}\mathbf{\vc{\lambda}_7}} = \frac{1}{5}\left(\;0,0,0,{\color{clr1}\mathbf{2}},0,{\color{clr3}\mathbf{3}},\ldots\;\right)$\\
			\end{tabular}
		}\\
		Update: & $\vc{m}=\left(\;0,0,0,{\color{clr10}\mathbf{0}},0,{\color{clr10}\mathbf{0}},{\color{clr10}\mathbf{1}},\ldots\;\right)$\vspace{0.5mm}\\
	\hline\\[-2.5ex]
			Result: & 
			{\tabcolsep0pt
				\begin{tabular}{l}
				$\vc{m} = \left(\;0,0,0,0,0,0,1\ldots\;\right)$\\
			  $\Lambda = \left[\;\unit{1},\unit{2},\unit{3},\unit{4},\vc{\lambda}_5,\vc{\lambda}_6,\vc{\lambda}_7\;\right]$.
				\end{tabular}
			}
		\end{tabular}
		\end{minipage}
	}
	\vspace{-1mm}
\caption{\label{fig:factorization}Factorization of a mixture $\vc{m}$, $\|\vc{m}\|_0=4$. In each step, two non-zero positions in $\vc{m}$ are selected, a new linear interpolation $\lambda$ is added to the mixture list $\Lambda$, and $\vc{m}$ is updated ($\ldots$ denote \hl{trailing} zeros).}
\vspace{-4mm}
\end{figure}
Finally, we update $\vc{m}$,\vspace{-1mm}
\begin{align}
	\vc{m} &\mapsfrom \vc{m} - \comp{m}{i}\unit{i} - \comp{m}{j}\unit{j} + \left(\comp{m}{i}+\comp{m}{j}\right)\unit{N+1}.
\end{align}
This update removes one non-zero entry from $\vc{m}$ in total, and creates a linear interpolation $\vc{\lambda}_{N+1}$.
Here, we used the embedding of mixtures in $\R^\infty$ to add mixtures that, informally speaking, use positions ``behind'' those dimensions of $\R^\infty$ that previously carried information. \hl{We therefore use the embedding in $\R^\infty$ to the same effect described in Hilbert's Grand Hotel thought experiment~{\cite{Hilbert}}, in which a fully booked hotel with infinitely many rooms can always accommodate a countable, potentially infinite number of new guests. Unlike the original thought experiment, in which guests move rooms to free up the room with the smallest index, we use the embedding to append mixtures with larger and larger $\maxc{m}$ to a countable set of mixtures with finite norms which, in the limit, may span $\R^\infty$.}

When repeated until $\|\vc{m}\|_0=1$, we obtain a series of linear interpolations $\vc{\lambda}_i$ with maximum non-zero positions\vspace{-1mm}
\begin{align}
  \widehat{k}\left(\vc{\lambda}_i\right) < i,
\end{align}
which is a direct consequence of ``appending'' linear interpolations to previously used dimensions in $\R^\infty$. Figure~\ref{fig:factorization} shows an example for the factorization of mixtures.\\

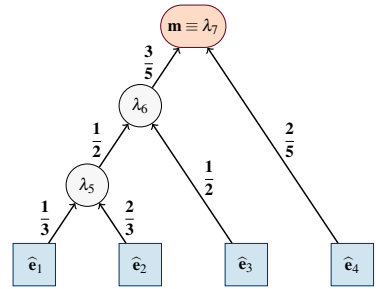
\begin{figure}[b]
\centering
\resizebox{0.55\columnwidth}{!}{%
\begin{tikzpicture}
\node[shape=rectangle,draw=clr11,fill=clr7,minimum size=0.8cm] (1) at (0,0) {~~$\unit{1}$~~};
\node[shape=rectangle,draw=clr11,fill=clr7,minimum size=0.8cm] (2) at(2,0) {$\unit{2}$};
\node[shape=rectangle,draw=clr11,fill=clr7,minimum size=0.8cm] (3) at (4,0) {$\unit{3}$};
\node[shape=rectangle,draw=clr11,fill=clr7,minimum size=0.8cm] (4) at(6,0) {$\unit{4}$};

\node[shape=circle,draw=black,fill=clr6,minimum size=0.8cm] (5) at (1,1.5) {$\vc{\lambda}_5$};
\node[shape=circle,draw=black,fill=clr6,minimum size=0.8cm] (6) at (2,3) {$\vc{\lambda}_6$};
\node[shape=rounded rectangle,draw=clr1,fill=clr5,minimum size = 0.8cm] (7) at (3,4.5) {$\vc{m}\equiv\vc{\lambda}_7$};

\path[->, thick] (1) edge node[left] {\raisebox{12pt}{$\mathbf{\dfrac{1}{3}~}$}} (5);
\path[->, thick] (2) edge node[right] {\raisebox{12pt}{$\mathbf{~\dfrac{2}{3}}$}} (5);
\path[->, thick] (3) edge node[right] {\raisebox{12pt}{$\mathbf{~\dfrac{1}{2}}$}} (6);
\path[->, thick] (4) edge node[right] {\raisebox{12pt}{$\mathbf{~\dfrac{2}{5}}$}} (7);
\path[->, thick] (5) edge node[left] {\raisebox{12pt}{$\mathbf{\dfrac{1}{2}}~$}} (6);
\path[->, thick] (6) edge node[left]  {\raisebox{12pt}{$\mathbf{\dfrac{3}{5}}~$}} (7);
\end{tikzpicture}
}
\vspace{-1mm}
\caption{\label{fig:tree}Factorization tree of the example in Fig.~\ref{fig:factorization}. Leaf nodes (boxes) store unit vectors representing the input segment IDs, internal nodes (circles) perform linear interpolations between two children, and the root (rounded rectangle) corresponds to the original mixture $\vc{m}$.}
\end{figure}
As depicted in Fig.~\ref{fig:tree}, such a series of linear interpolations can be represented as a binary tree with edge weights. Leaf nodes represent the input volume's segments and the root represents the final mixture $\vc{m}$. Each internal node represents a linear interpolation $\vc{\lambda}$ with exactly two children corresponding to the two non-zero entries $i,j$ in $\vc{\lambda}$. Edge weights are given by $\comp{\lambda}{i}$ and $\comp{\lambda}{j}$.\\

Carrying out the factorization for a set of mixtures while re-using identical nodes results in a directed acyclic graph (DAG) $\G = \left(N,E,W\right)$ \hl{as} depicted in Figure~\ref{fig:DAG}. The node set $N$ contains nodes with an in-degree of $0$ (sources) representing the original segment IDs, internal nodes with an in-degree of $2$, and sinks with an out-degree of $0$. The final mixture for each voxel is represented by either sources or sinks. We define the edge direction of $(i,j)\in E$ as ``from $i$ to $j$'', with the notion of mixture $i$ ``contributes to'' mixture $j$:\vspace{-1mm}
\begin{align}
	\exists \left(i,j\right)\in E:\, w_{ij}\neq 0\in W \;\;\Leftrightarrow\;\; \exists \vc{\lambda_j}\in\Lambda,\,i\in\N_0\,:\, \comp{\lambda_j}{i}=w_{ij}.
\end{align}
Since each node except for sources has an in-degree of exactly two, we store the connectivity information for each node as incoming edges. 
For each edge $\left(i,j\right)$, we call node $i$ a \textsl{predecessor} of $j$ and, conversely, $j$ a \textsl{successor} of $i$.

\subsection{Compression}
\label{sec:compression}
\begin{figure}[t]
\centering
\resizebox{0.55\columnwidth}{!}{%
\begin{tikzpicture}
\node[shape=rectangle,draw=clr11,fill=clr7,minimum size=0.8cm] (1) at (0,0) {~~$\unit{1}$~~};
\node[shape=rectangle,draw=clr11,fill=clr7,minimum size=0.8cm] (2) at(2,0) {$\unit{2}$};
\node[shape=rectangle,draw=clr11,fill=clr7,minimum size=0.8cm] (3) at (4,0) {$\unit{3}$};
\node[shape=rectangle,draw=clr11,fill=clr7,minimum size=0.8cm] (4) at(6,0) {$\unit{4}$};

\node[shape=circle,draw=black,fill=clr6,minimum size=0.8cm] (5) at (1,1.5) {$\vc{\lambda}_5$};
\node[shape=circle,draw=black,fill=clr6,minimum size=0.8cm] (6) at (2,3) {$\vc{\lambda}_6$};
\node[shape=rounded rectangle,draw=clr1,fill=clr5,minimum size = 0.8cm] (7) at (3,4.5) {$\vc{m}\equiv\vc{\lambda}_7$};

\node[shape=circle,draw=black,fill=clr6,minimum size=0.8cm] (8) at (5,-1.5) {$\vc{\lambda}_8$};
\node[shape=rounded rectangle,draw=clr1,fill=clr5,minimum size = 0.8cm] (9) at (3,-2.4) {$\vc{m^\prime}\equiv\vc{\lambda}_9$};

\path[->, thick] (1) edge node[left] {\raisebox{12pt}{$\mathbf{\dfrac{1}{3}~}$}} (5);
\path[->, thick] (2) edge node[right] {\raisebox{12pt}{$\mathbf{~\dfrac{2}{3}}$}} (5);
\path[->, thick] (3) edge node[right] {\raisebox{12pt}{$\mathbf{~\dfrac{1}{2}}$}} (6);
\path[->, thick] (4) edge node[right] {\raisebox{12pt}{$\mathbf{~\dfrac{2}{5}}$}} (7);
\path[->, thick] (5) edge node[left] {\raisebox{12pt}{$\mathbf{\dfrac{1}{2}}~$}} (6);
\path[->, thick] (6) edge node[left]  {\raisebox{12pt}{$\mathbf{\dfrac{3}{5}}~$}} (7);
\path[->, thick] (3) edge node[left] {\raisebox{-18pt}{$\mathbf{\dfrac{1}{4}}~$}} (8);
\path[->, thick] (4) edge node[right] {\raisebox{-18pt}{$~\mathbf{\dfrac{3}{4}}$}} (8);
\path[->, thick] (8) edge[bend left] node[right] {\raisebox{-24pt}{$~\mathbf{\dfrac{2}{5}}$}} (9);
\path[->, thick] (5) edge[bend right] node[left] {\raisebox{-24pt}{$~\mathbf{\dfrac{3}{5}}$}} (9);
\end{tikzpicture}
}
\vspace{-2mm}
\caption{\label{fig:DAG}Example directed acyclic graph (DAG) corresponding to the factorization of $\vc{m}=\left(\;0.1,0.2,0.3,0.4,\ldots\;\right)$ and $\vc{m^\prime}=\left(\;0.2,0.4,0.1,0.3,\ldots\;\right)$. Mixture $\vc{\lambda}_5$ can be re-used, since the ratio between the first two components is the same ($0.1:0.2\;\equiv\;0.2:0.4$).}
\vspace{-4mm}
\end{figure}
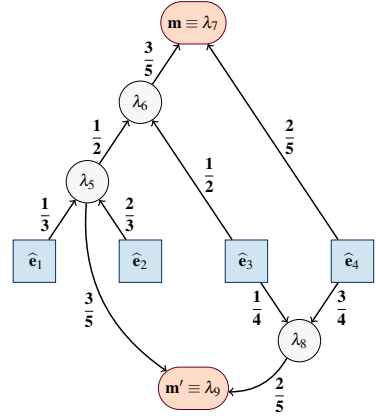

From the previous section, it is intrinsically clear that the factorization into linear interpolations is not unique: in each step \textsl{any} two non-zero elements of $\vc{m}$ can be picked. Our compression method exploits this degree of freedom to generate as many redundant nodes as possible. For instance, Fig.~\ref{fig:DAG} shows that $\lambda_5$ can be re-used in the factorization of both $\vc{m}$ and $\vc{m^\prime}$, since they both mix $\unit{1}$ and $\unit{2}$ in the same ratio $1:2$, albeit with different weights of $0.5$ and $0.6$.\\

Finding the factor with the highest re-usability is a hard problem. In the first step, we have \hl{$N$ choose 2} possible picks $i,j$, where $N$ is the number of segments in the input. 
Picking $i,j$ removes this combination from subsequent picks. However, we add a new option to pick from \hl{($N-1$ choose 2)}. To maximize overall re-use of nodes, we thus cannot process the factorization steps independently of one another. The full search space offers a total of $P$ choices, with\vspace{-0.5mm}
\begin{empheq}[box=\hlm]{align}
P&=\prod_{i=2}^{N} \genfrac(){0pt}{0}{\,i\,}{\,2\,} = \frac{1}{2}\prod_{i=2}^{N} i\left(i-1\right) =\frac{1}{2N}\left(N!\right)^2.
\end{empheq}
The problem is further exacerbated by the sheer \hl{number} of mixtures to be considered and the fact that only nodes with the same ratio between components $i$ and $j$ can be re-used. Our method therefore relies on a greedy algorithm to find a candidate pick $i,j$ that has a good re-use probability.
We consider two greedy strategies. The first strategy, which we call \textbf{max-occurrence}, picks the pair of indices $i,j$ that most frequently corresponds to non-zero components of mixtures in $\His$. The rationale behind this strategy is that, even though we are ultimately interested in recycling nodes representing mixtures $\vc{\lambda} =  (1-w)\unit{i} + w\unit{j}$, frequent combinations of non-zero indices $i,j$ are good candidates. The reason is that $w\in(0,1)$ can be quantized to increase re-use.
Formally, we define a counting function $\phi_1$,\vspace{-0.5mm}
\begin{align}
	\mathcal{C} &:= \left\{i,j\in\left\{1,\ldots,N\right\}:i\neq j\right\}\nonumber\\
	\phi_1 &:\hphantom{=} \mathcal{C} \rightarrow \N_0\nonumber\\
	\phi_1(i,j) &:= \left|\left\{\vc{h}\in\His : \comp{h}{i}\neq 0 \wedge \comp{h}{j}\neq 0\right\}\right|,
\end{align}
and compute\vspace{-1mm}
\begin{align}
	\label{eq:greedy1}
	i,j &=\argmax_{k,l\in\mathcal{C}}\; \phi_1(k,l).
\end{align}
We call our second strategy \textbf{max-reduction}. It exploits that after factoring out a mixture $\lambda$ in $\vc{m}$ with $\|\lambda\|_0=2$ and $\|\vc{m}\|_0=N$, a mixture $\vc{m^\prime}$ with $\|\vc{m^\prime}\|_0=N-1$ remains. Thus, picking $i,j$ reduces our choices from \hl{$N$ choose 2} to \hl{$N-1$ choose 2}, a reduction by $N-1$ if $N>2$ and by $2$ if $N=2$. Since this decrease of choices is directly equivalent to shrinking the search space, our strategy is to find a pair $i,j$ minimizing the size of the remaining search space.
Formally, we define a second counting function $\phi_2$,\vspace{-1mm}
\begin{align}	
	\phi_2 &:\hphantom{=}\mathcal{C}\rightarrow\N_0\nonumber\\
	\phi_2(i,j) &:= \sum_{\vc{h}\in\His}
		\begin{cases}
			0 &\mbox{if }\comp{h}{i}=0 \vee \comp{h}{j}=0\\
			\|\vc{h}\|_0-1 &\mbox{else if } \|\vc{h}\|_0>2\\
			2 &\mbox{else if } \|\vc{h}\|_0=2
		\end{cases}\;,
\end{align}
and compute\vspace{-1mm}
\begin{align}
\label{eq:greedy2}
	i,j &= \argmax_{k,l\in\mathcal{C}}\;\phi_2(k,l).
\end{align}
These two strategies are compared in the Results section. Once we have found the best pair $i,j$ according to one of the strategies, we proceed by factoring all occurrences of $i,j$. 
We repeat this process until $\His$ is fully factorized, i.e., $\forall \vc{h}\in\His: \|\vc{h}\|_0=1$. To improve re-use of the nodes thus generated, we use scalar quantization on the ratios of the linear interpolations. Note in this context that linear interpolation ratios cannot be $0$ or $1$, since this would exactly replicate a previous mixture. Consequently, our scalar quantizer partitions the open range $(0,1)$ into $2^b$ bins, where $b$ is the bitrate of the quantization. Each bin is represented by a single floating point codeword that best represents the weights in the bin. In addition, we store an empirical standard deviation $\sigma$ for each bin to facilitate estimating the error of the quantization throughout the entire reconstruction process.
Since the diversity of interpolation weights is greatly reduced by the quantization step, more nodes in the Mixture Graph become identical and can be re-used.\\
\begin{table}[b]
\captionsetup{skip=0pt}
\caption{\label{tab:bitstream}Overview of the output bitstream.}
	\renewcommand{\arraystretch}{1.1}
	\tabcolsep3pt
	\centering
	\begin{tabular}{ll|l|l}
	\\[-1.8ex]
		ID & description     & \# bits & range/comments\\
		\hline
		(1) & \# sources 			  & 32bits & $[0,\ldots,2^{32}-1]$\\
		(2) & fractional input ?  &  ~~1bit & $\exists\vc{h}\in\His(\cdot,0):\;\|\vc{h}\|_0>1$ ?\\
		(3) & bits per node   &  ~~6bits & $[1,\ldots,64]$\\
		(4) & $\sigma_{\max}$ & 32bits & max. $\sigma$ of quantization\\
		(5) & 3D resolution		& 3$\times$32bits & resolution of $\His(\cdot,0)$\\
		(6) & quantization bitrate & ~~4bits & $[1,\ldots,16]$\\
		(7) & \# internal nodes & var bits & \# bits defined at (3)\\
		\hline\\[-1.8ex]
		\multicolumn{4}{l}{For each level in $\His$}\\
		\hline
		(8) & \# voxel bits & var bits & \# bits defined at (3)\\
		(9) & $i_{\min}(l)$ & var bits & \# bits defined at (3)\\
		\hline\\[-1.8ex]
		\multicolumn{4}{l}{Bulk data}\\
		\hline
		(A) & \multicolumn{2}{l|}{one mixture ID per voxel} & \# bits defined at (8)\\
		(B) & \multicolumn{2}{l|}{description of mixtures} & \# bits = $2\times$(3) + (6)\\
		(C) & \multicolumn{2}{l|}{scalar quantizer codebook} & 20bits for each weight and $\sigma$\\
	\end{tabular}
\end{table}

The final output of the compression step is a bitstream storing the compressed Mixture Graph. Sources are not stored explicitly since they correspond to input segmentation IDs. The number of sources, however, is stored in the bitstream's header, along with the volume's dimensions, a flag indicating whether the input volume contained a fractional segmentation (i.e., input segments are already mixtures), the bitrate of the scalar quantizer, etc. Table~\ref{tab:bitstream} provides an overview of the output format. For each node, we store a triple of values consisting of two node IDs and one index into the scalar quantizer's codebook of interpolation weights. This is done at a fixed bitrate of $2\lceil\log_2\left(\#\mathrm{nodes}\right)\rceil+b$. For each level of $\His$, we compute the minimum and maximum ID of all nodes referenced in this level, i.e.,
\begin{align}
	i_{\min}(l) &:= n : \comp{h}{n}\neq 0 \wedge \comp{h}{k}=0\quad\forall \vc{h}\in\His, k<n,\nonumber\\
	i_{\max}(l) &:= \max_{\vc{h}\in\His(\cdot,l)} \widehat{k}_{\vc{h}}.
\end{align}
We store $i_{\min}(l)$ in the bitstream's header for each $l$. For each voxel, we then store its node ID minus the minimum node ID of that level, again, using a fixed bitrate of\vspace{-1mm}
\begin{align}
	r(l)&=\left\lceil\log_2\left(i_{\max}(l)-i_{\min}(l)+1\right)\right\rceil
\end{align}
for each voxel in this level. Finally, we store the scalar quantizer's code values and standard deviations in 20 bits each.

\subsection{Reconstruction}
\label{sec:reconstruct}
\begin{figure}[b]
\vspace{-2mm}
\centering
\resizebox{0.85\columnwidth}{!}{%
\begin{tikzpicture}
\node[shape=rectangle,draw=clr11,fill=clr7,minimum size=0.8cm] (1) at (0,0) {$0$};
\node[shape=rectangle,draw=clr11,fill=clr7,minimum size=0.8cm] (2) at (2,0) {$0$};
\node[shape=rectangle,draw=clr11,fill=clr7,minimum size=0.8cm] (3) at (4,0) {$0$};
\node[shape=rectangle,draw=clr11,fill=clr7,minimum size=0.8cm] (4) at (6,0) {$0$};
\node[shape=rectangle,draw=clr11,fill=clr7,minimum size=0.8cm] (5) at (8,0) {$0$};

\node[shape=circle,draw=black,fill=clr4,minimum size=0.8cm] (6) at (1,1.5)  {$1$};
\node[shape=circle,draw=black,fill=clr7!50!clr8,minimum size=0.8cm] (7) at (7,1)  {$1$};

\node[shape=circle,draw=black,fill=clr8,minimum size=0.8cm] (9) at (6,2)  {$2$};

\node[shape=rounded rectangle,draw=black,fill=clr8!50!clr9,minimum size = 0.8cm] (10) at (4,3) {$3$};

\path[->, thick] (1) edge node[left] {} (6);
\path[->, thick] (2) edge node[right] {} (6);

\path[->, thick] (4) edge node[right] {} (7);
\path[->, thick] (5) edge node[right] {} (7);

\path[->, thick] (3) edge node[left] {} (9);
\path[->, thick] (7) edge node[left]  {} (9);

\path[->, thick] (6) edge node[left] {} (10);
\path[->, thick] (9) edge node[right] {} (10);

\end{tikzpicture}
}
\caption{\label{fig:dagtopo}Topological sorting of the DAG. Node numbers indicate the distance to the closest source. During parallel reconstruction smaller numbers have to be processed before larger numbers. The largest number corresponds sto the number of synchronization steps required. Here, the orange node has only successors with a value of 3. It can be processed together with nodes labelled '$1$' or with nodes labelled '$2$'.}
\end{figure}
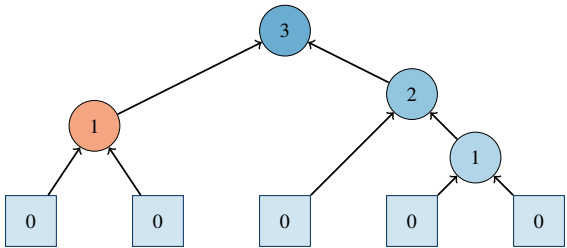

In order to reconstruct the original normalized histogram mipmap, we first perform a topological sorting of the nodes $\Lambda$ in the Mixture Graph. Despite the factorization step resulting in a \hl{``soft''} topological order (children-before-parents), this step is necessary in order to establish synchronization points for parallel reconstruction. 
Topological sorting resolves dependencies when reconstructing the original mixtures (also see Fig.~\ref{fig:dagtopo}). We begin by assigning a value of $0$ to each source in the graph. Each remaining node with predecessor values $v_1,v_2$ is assigned a value of $\max\left(v_1,v_2\right)+1$. We call these values $v_i$ the \textsl{topological level} of node $i$ and $\max_{v_i}$ the topological depth of the graph. Exploiting the children-before-parents relation described above, this amounts to a linear scan of all nodes.

On parallel architectures, such as GPUs, nodes with identical values can be processed without synchronization. Conversely, this also means that the topological depth is equivalent to the required number of synchronization barriers. Furthermore, since we only need to ascertain that all nodes with smaller values are processed before the first node with a larger value, we can defer evaluating certain nodes. Consider a node with topological level $v$ and a successor level of $v_s$. If $v_s-v>1$, then the node can be evaluated in parallel together with node levels $v,v+1,\ldots,v_s-1$. Since in our experience much more nodes are generated with small values than with large ones, this can be used for load-balancing. We start by processing all nodes with a value of 0 and a successor value of 1. We then continue processing nodes with value of 0 but a successor value of 2, until we have processed at least $\lceil N/p\rceil$ nodes, where $N$ is the total \hl{number} of nodes and $p$ is the topological depth of the DAG. We then repeat this procedure for the next level, until we reached the sinks in the DAG.\\

The Mixture Graph not only stores a factorization of the mixtures in $\His$, but also represents the computations necessary to reconstruct mixtures of segment attributes at all scales of a mipmap. Therefore, it is also possible to assign each segment an RGB$\alpha$ color and rebuild the resulting RGB$\alpha$ mipmap efficiently. Traditionally, this requires to assign colors first and then to rebuild the mipmap from scratch. Using the Mixture Graph, only all potentially unique RGB$\alpha$ colors in the resulting mipmap are computed by propagating colors through the Mixture Graph. The mipmap itself is essentially a paletted texture. In the Results section we show that the 1,046 megavoxel Hippocampus data set with 353 segments and a mipmap comprising more than 1,195 \hl{million} voxels can be factored into a little more than 1 million nodes.
Consequently, instead of re-computing the color values of 1,195 \hl{million} voxels, we only need to update 1 million colors by linear interpolations. \hl{A partial update of the transfer function would trigger a ``push'' scattering propagation of through the graph.} Since scattering is much harder to parallelize than gathering, our system only performs full updates using parallel gathering. As we show in the Results section, the performance of such updates is still well within real-time requirements. Note that, apart from other benefits, assigning an RGB$\alpha$ value to each voxel of the aforementioned data set would amount to more than 4.6GB, whereas our data representation requires only 1.36GB.

Finally, we would like to point out that, as long as the input data contains only one segment ID per voxel, we can always reconstruct the lowest level (input resolution) without loss from the Mixture Graph. \hl{Higher-level mixtures are encoded with a small loss, 
and we would like to refer the reader to the Results section for details.}

\subsection{Error Propagation} 
\label{sec:error}
To estimate the error in the reconstruction, the scalar quantizer stores the empiric standard deviation in addition to the code value per bin. The quantization error can be propagated following the standard rules for error propagation during the reconstruction:\vspace{-1mm}
\begin{align}
	Q&=a+b\quad \rightarrow \delta Q = \sqrt{\left(\delta a\right)^2 + \left(\delta b\right)^2}\nonumber\\
	Q&=a\times b\quad \rightarrow \delta Q = \sqrt{Q^2\left(\left(\frac{\delta a}{a}\right)^2 + \left(\frac{\delta b}{b}\right)^2\right)}.
\end{align}
Sources in the Mixture Graph represent certain input segment IDs $\unit{i}, \delta\unit{i} = 0$. Successive linear interpolations in the reconstruction using uncertain weights introduce uncertainties into the mixtures of the input segments. In particular, we consider the linear interpolation between two potentially uncertain vectors $\vc{a}\pm\delta \vc{a}$, $\vc{b}\pm\delta \vc{b}$ using an uncertain weight $w\pm\delta w$. Since $\vc{a},\vc{b}$ model mixtures of segments, we treat the error of each component $i$ in these vectors separately. We obtain\vspace{-1mm}
\begin{align}
	\label{eq:error}
	\comp{\mu}{i} &= (1-w)\comp{a}{i}+w\comp{b}{i}\nonumber\\
	\comp{\delta\mu}{i} &=
			\sqrt{\comp{\mu}{i}^2\left(
			\left(\frac{\comp{\delta a}{i}}{\comp{a}{i}}\right)^2 + 
			\left(\frac{\comp{\delta b}{i}}{\comp{b}{i}}\right)^2 +
			2\left(\frac{\delta w}{w}\right)^2\right)}.
\end{align}
\hl{Despite the fact that segments are typically spatially coherent, it is worth noting that}  the above independent treatment of the error propagation is correct since the uncertain weights are independent of the mixture vectors. We also validated this experimentally by sampling weights from standard distributions during the reconstruction. The above error propagation equations provide us with the means to quantify the error of volume region queries. 

\subsection{Fast Sub-Volume Queries}
\label{sec:svq}
\begin{figure}[tb]
\centering

\begin{tikzpicture}[scale=0.65, every node/.style={scale=0.65}]

	\begin{scope}[shift={(5.2cm,0.0cm)}]
		\draw[step=0.4cm,gray,very thin] (0.6,-0.2) grid(4.4,5.4);

		\filldraw[fill=clr9] (1.2,-0.4) rectangle (1.6,0);
		\filldraw[fill=clr7] (1.6,-0.4) rectangle (3.2,0);
		\filldraw[fill=clr8] (3.2,-0.4) rectangle (4,0);
		\begin{scope}[shift={(0.2cm,0)}]
			\foreach \x in {2,3,4,5,6,7,8,9,10} \draw (\x*0.4 cm,0) node[anchor=north] {$\x$};
		\end{scope}

		\filldraw[fill=clr9] (0.4,1.2) rectangle (0.8,1.6);
		\filldraw[fill=clr7] (0.4,1.6) rectangle (0.8,3.2);
		\filldraw[fill=clr7] (0.4,3.2) rectangle (0.8,4.8);
		\begin{scope}[shift={(0.8,0.2cm)}]
			\foreach \y in {0,1,2,3,4,5,6,7,8,9,10,11,12} \draw(0.08,\y*0.4 cm) node[anchor=east] {$\y$};
		\end{scope}

		\draw[black,thin] (0.6,0) -- (4.6,0);
		\draw[black,thin] (0.8,-0.2) -- (0.8,5.4);

		\begin{scope}[shift={(1.2,0.4)}]
			\foreach \x in {0,1,2,3,4,5,6} \filldraw[fill=clr9] (\x*0.4,0.8) rectangle(\x*0.4 + 0.4,1.2);
		\end{scope}

		\begin{scope}[shift={(1.2,1.6)}]
			\foreach \y in {0,1,2,3,4,5,6,7} \filldraw[fill=clr9] (0,\y*0.4) rectangle (0.4,\y*0.4 + 0.4);
		\end{scope}
	
		\begin{scope}[shift={(1.6,1.6)}]
			\foreach \y in {0,1} {
				\filldraw[fill=clr7] (0,\y*1.6) rectangle(1.6,\y*1.6+1.6);
				\draw[step=0.4cm,clr6,very thin] (0.02,\y*1.6+0.02) grid (1.58,\y*1.6+1.58);
			}
		\end{scope}

		\begin{scope}[shift={(3.2,1.6)}]
			\foreach \y in {0,1,2,3} {
				\filldraw[fill=clr8] (0,\y*0.8) rectangle(0.8,\y*0.8+0.8);
				\draw[step=0.4cm,clr6,very thin] (0.02,\y*0.8+0.02) grid (0.78,\y*0.8+0.78);
			}
		\end{scope}
	\end{scope} 

\draw[step=0.4cm,gray,very thin] (-0.2,-0.2) grid(5,5.4);

\foreach \x in {0,1,2,4,5,6,7,8,9,10,11} \filldraw[fill=clr6] (\x*0.4,-1.7) rectangle (\x*0.4+0.4,-1.4);
\foreach \x in {0,1,2,3,5} \filldraw[fill=clr6] (\x*0.8,-1.3) rectangle (\x*0.8+0.8,-1);
\foreach \x in {0,2} \filldraw[fill=clr6] (\x*1.6,-0.9) rectangle (\x*1.6+1.6,-0.6);
\filldraw[fill=clr9] (1.2,-1.7) rectangle (1.6,-1.4);
\filldraw[fill=clr8] (3.2,-1.3) rectangle (4,-1);
\filldraw[fill=clr7] (1.6,-0.9) rectangle (3.2,-0.6);
\draw (-0.1,-1.55) node[anchor=east] {$0$};
\draw (-0.1,-1.15) node[anchor=east] {$1$};
\draw (-0.1,-0.75) node[anchor=east] {$2$};
\filldraw[fill=clr9] (1.2,-0.4) rectangle (1.6,0);
\filldraw[fill=clr7] (1.6,-0.4) rectangle (3.2,0);
\filldraw[fill=clr8] (3.2,-0.4) rectangle (4,0);
\begin{scope}[shift={(0.2cm,0)}]
	\foreach \x in {0,1,2,3,4,5,6,7,8,9,10,11} \draw (\x*0.4 cm,0) node[anchor=north] {$\x$};
\end{scope}
\draw[decorate,decoration={brace,amplitude=4pt},xshift=0,yshift=0] (-0.4,-1.7) -- (-0.4,-0.6) node [black,midway,xshift=-0.4cm] {$l_x$};

\draw[black,thin] (-0.2,0) -- (5,0);
\draw[black,thin] (0,-0.2) -- (0,5.4);

\filldraw[fill=clr6] (-0.9,0) rectangle (-0.6,1.6);

\foreach \y in {1,2} \filldraw[fill=clr7] (-0.9,\y*1.6) rectangle (-0.6,\y*1.6+1.6);
\draw (-0.75,0) node[anchor=north] {$2$};
\foreach \y in {0,1,2,3,4,5} \filldraw[fill=clr6] (-1.3,\y*0.8) rectangle (-1,\y*0.8+0.8);
\draw (-1.15,0) node[anchor=north] {$1$};
\foreach \y in {0,1,2,4,5,6,7,8,9,10,11,12} \filldraw[fill=clr6] (-1.7,\y*0.4) rectangle (-1.4,\y*0.4+0.4);
\draw (-1.55,0) node[anchor=north] {$0$};
\filldraw[fill=clr9] (-1.7,1.2) rectangle (-1.4,1.6);
\draw[decorate,decoration={brace,amplitude=4pt,mirror},xshift=0,yshift=0] (-1.7,-0.35) -- (-0.6,-0.35) node [black,midway,yshift=-0.4cm] {$l_y$};

\filldraw[fill=clr6] (-0.9,4.8) rectangle (-0.6,5.42);
\filldraw[fill=clr6] (-1.3,4.8) rectangle (-1,5.42);
\fill[fill=white] (-1.72,5.35) rectangle (-0.58,5.45);

\filldraw[fill=clr9] (-0.4,1.2) rectangle (0,1.6);
\filldraw[fill=clr7] (-0.4,1.6) rectangle (0,3.2);
\filldraw[fill=clr7] (-0.4,3.2) rectangle (0,4.8);
\begin{scope}[shift={(0,0.2cm)}]
	\foreach \y in {0,1,2,3,4,5,6,7,8,9,10,11,12} \draw(0.08,\y*0.4 cm) node[anchor=east] {$\y$};
\end{scope}

\begin{scope}[shift={(1.2,1.2)}]
	\filldraw[fill=clr6] (0,0) rectangle (0.4,0.4);
	\filldraw[fill=clr6] (0.4,0) rectangle (2.0,0.4);
	\filldraw[fill=clr6] (2.0,0) rectangle (2.8,0.4);
	
	\filldraw[fill=clr6] (0,0.4) rectangle (0.4,2.0);
	\filldraw[fill=clr6] (0,2.0) rectangle (0.4,3.6);
	\filldraw[fill=clr6] (0.4,0.4) rectangle (2.0,2.0);
	\filldraw[fill=clr6] (0.4,2.0) rectangle (2.0,3.6);
	\filldraw[fill=clr6] (2.0,0.4) rectangle(2.8,2.0);
	\filldraw[fill=clr6] (2.0,2.0) rectangle(2.8,3.6);
\end{scope}

\end{tikzpicture}
\caption{\label{fig:footprint}Footprint assembly in 2D. \textbf{Left:} The 2D footprint formed by the tensor product of 1D footprints consists of rectangles. Alignments and extents of samples of the mimmap are shown next to each axis. \textbf{Right}: Rectangles in the 2d footprint are tesselated using squares matching the smaller side of each rectangle. Each square corresponds to one sample from the normalized histogram mipmap.}
\vspace{-4mm}
\end{figure}
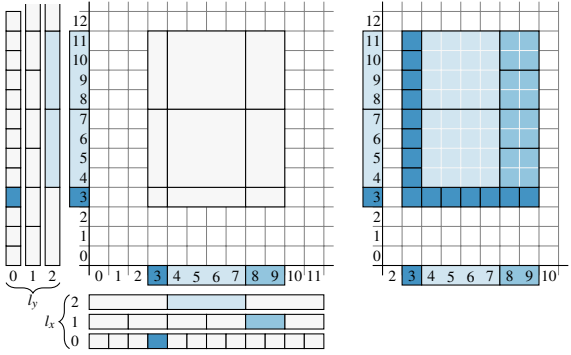

To compute the histogram over a sub-volume $\left[x,x+\Delta x\right]\times\left[y,y+\Delta y\right]\times\left[z,z+\Delta z\right]$, we follow a footprint assembly~\cite{texram,patent} approach. First, we decompose the 1D ranges along $x,y$ and $z$ according to their alignment with the mip levels in the normalized histogram mipmap $\His$. Then, the tensor product of these ranges is comprised of rectangular boxes as depicted in Fig.~\ref{fig:footprint}. Each of these boxes is tesselated by cubes of a side length $l$ equal to the shortest of the three box extents. Each cube represents a single sample from $\His$ at level $\log_2(l)$. The contribution of each sample to the final result is denormalized (multiplication by $l^3$) before being accumulated. In order to evaluate samples from $\His$ efficiently, a lookup table is pre-computed on the CPU that stores one mixture per node. We then use error propagation (Eq.~\ref{eq:error}) to track the uncertainty of the interpolation weights through the lookup table. This allows us to predict the error made in range queries. Typical errors using 512 bins for quantization are about 0.3\%, with the maximum error for a segment not exceeding 0.5\%. As a direct benefit of the Mixture Graph only considering convex combinations of nodes, the sum of the expectations always sum up exactly to the volume of the query region. Our error estimate is thus unbiased.

The task of querying ranges of the segmented volume is not well suited for the GPU, since the number of non-zero entries in the sparse mixtures is hard to predict. While it is generally possible to na\"{i}vely reconstruct a dense histogram, a dense approach is unlikely to scale well for volumes with a high number of segment IDs, since the sparsity in the range is not used. Furthermore, while we only discussed axis-aligned sub-volumes in this section, footprint assembly can also be performed for ``oddly-shaped'' regions---albeit at a substantially higher computational cost.

\subsection{Implementation Details}
\label{sec:impl}
\noindent\textbf{Maximum Search.} Our compression algorithm relies on quickly finding the best pair of indices $i,j$ with respect to one of the greedy strategies defined in Eq.~\eqref{eq:greedy1} and Eq.~\eqref{eq:greedy2}. We note that at first the best and second-best pair tend to be well-separated in terms of their score $\phi_1$ or $\phi_2$. In later stages, progress slows down as choices become more ambiguous. We therefore introduce a threshold $\tau$. In the first stage, we factorize all mixtures $\vc{m}$, $\|\vc{m}\|_0>\tau$, skipping over many entries in lower levels of the mipmap. Once we have successfully exhausted options in this search space, we factorize the remaining mixtures with only $\tau$ or less components in scanline order. Using a cutoff $\tau=3$, about half of all mixtures can be exempted in the first stage, resulting in about $2.73\times$ speed increase when compared to a na\"{i}ve exhaustive search when $\phi_1$ is used as a scoring function, and in about $1.79\times$ improvement for $\phi_2$. The reason is that $\phi_2$ reduces the search space quicker and results in a better performance overall.\\

\noindent\textbf{Interpolation Symmetries.} To find node redundancies, we consider triples $(i,j,w)$ to denote linear interpolations between segments $s_i$ and $s_j$: $(1-w)s_i + w s_j$. Thus, each triple $(i,j,w)$ is equivalent to $(j,i,1-w)$. In our implementation, we use this symmetry to improve the accuracy of the quantized weight. To do so, we quantize both $w$ and $1-w$ using the codebook generated by the scalar quantizer and keep the one resulting in the lower error. We then store either $(i,j,w)$ or $(j,i,1-w)$ in the output stream.\\

\noindent\textbf{Sparse vectors over $\R^\infty$.} While it may seem appealing to implement the data structure for sparse vectors over $\R^\infty$ using a generic hash implementation (e.g., std::unordered\_map in C++), we found that the storage requirements of such implementations quickly become prohibitive. We therefore implemented the underlying data structure in C++ without the STL as a vector of pairs (index, value) that we keep sorted with respect to the index.

\section{Rendering}
To render the segmented volume directly from its Mixture Graph representation, we upload the entire binary representation to the GPU. Specifically, voxel data is stored in shader storage buffer objects (SSBOs); index structures such as node offsets, bit allocation counts etc.\ are stored as shader uniform constants. In addition, we allocate an SSBO with one vec4 color entry per node to hold the decoded color lookup table. The lookup table is updated with each transfer function update using a compute shader, as outlined before. Since our binary representation is accessed based on bit addresses, we use the uint64 data type provided to shaders by the GPU shader5 extension whenever necessary. For each voxel, we ultimately retrieve some index information (level node offset, bit allocation per voxel, etc.) plus the node \hl{ID} of the voxel. The node \hl{ID} is a direct reference into the color lookup table. In essence, rendering from the Mixture Graph is the same as rendering from a paletted texture, with the minor complications that the palette has to be reconstructed prior to rendering and address computations have to be performed in the shader.

While the Mixture Graph allows the efficient computation of any kind of numeric segmentation attribute, we shall for now assume that each segment is assigned an RGB$\alpha$ color through a transfer function. On the GPU, we traverse the Mixture Graph's nodes in topological order. For each node, we gather two colors and linearly interpolate between them using the quantized weight stored at the node. Once the palette is computed, rendering proceeds by identifying the address of any given voxel's raw data in a buffer containing the voxel IDs. We provide the number of levels as a uniform shader constant. Note that while the Mixture Graph stores a full mipmap and thus provides pre-filtered anti-aliasing, we still have to perform the computation of the mip level as well as the interpolation in the shader ``manually''. This is a drawback shared with virtually all modern paletted compression and rendering methods. To estimate the correct level of detail (LOD), we use the observation that the pixel coverage of a voxel is linear in viewspace depth $z$~\cite{Cohen98}. The corresponding equations can also be found in Section 8.14 of the OpenGL 4.5 specifications. We therefore upload viewport, camera parameters, volume dimensions as well as a uniform LOD bias to the shader. In addition, the step size of a volume raymarcher can be adjusted to the LOD, which requires opacity correction to be performed in case of semi-transparent structures.\\

\begin{figure}[b]
\vspace{-3mm}
\centering
	\includegraphics[width=0.49\columnwidth]{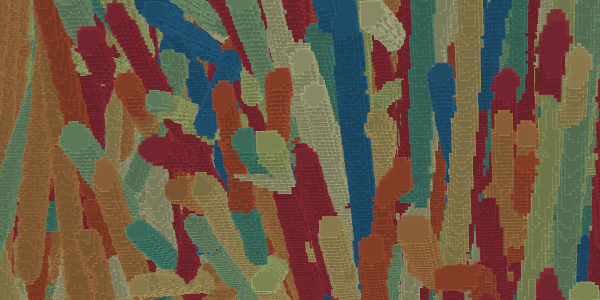}
	\includegraphics[width=0.49\columnwidth]{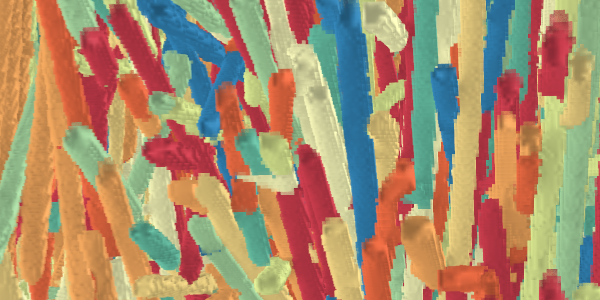}
	\vspace{-1mm}
	\caption{\label{fig:shading}Using on-the-fly opacity gradients with local ambient occlusion (left) vs.\ \hl{pre-computed} normals stored in a second Mixture Graph (right) for shading the Polymer data set.}
\end{figure}

\hl{\noindent \textbf{Empty Space Skipping.} The hierarchical nature of the Mixture Graph makes it possible to implement various empty space skipping schemes. Although a full evaluation of different empty space skipping methods is
beyond the scope of this paper, our current raymarcher implements a GPU-based scheme that: (i)
assigns opacity values of the current transfer function to the leafs, (ii) propagates these values through the Mixture Graph, and, (iii) uses the hierarchy for skipping longer segments. As long as the opacity value in parent voxels falls below a certain threshold, step (iii) traverses the hierarchy upwards to establish larger and larger blocks that can be skipped.
This simple yet flexible scheme is enabled by the Mixture Graph's ability to quickly recompute 
entire mipmaps. It results in an average 3$\times$ speedup when compared to the same implementation without mipmaps using a very low threshold of $10^{-8}$.} \\

\IGNORE{
Our renderer does not yet implement empty space skipping, since evaluating the trade-offs of different empty space skipping methods is beyond the scope of this paper. However, an easy way to add empty space skipping is to mark visible and non-visible segments with 0 and 1, respectively, and to propagate these values through the Mixture Graph. This results in a lookup table that will contain the percentage of visible voxels for each LOD voxel.\\
}
\hl{\noindent\textbf{Lighting.}} Since our input are volumes of categorical labels, the gradient of the volume cannot be used as normal for \hl{shading~{\cite{Lempitsky:10}}.} Instead, we estimate normals as follows. We start by tagging voxels at the boundary between two segments as \textit{features} and compute \hl{an} unsigned distance transform to these features using the method of Schneider et al.~\cite{Schneider:09:DT}. We erode our feature mask by two voxels and smooth the distance transform inside a thin band around the segment surfaces. This is necessary to compensate for artifacts due to the finite voxel resolution. We then compute gradients using centered differences, but the resulting gradients will be discontinuous across ridges of the medial axis. To alleviate this, we finally flip the feature mask and smooth the remaining gradients in the interior of the segments using isotropic diffusion. Once we have normal estimates, \hl{we quantize the normals~{\cite{Tarini:00:VQ}}} using vector quantization to 9 bit. To alleviate banding effects, we use 3D error diffusion~\cite{Lou:98:dithering}. This results in yet another volume of categorical values that we store in another Mixture Graph. Since the leafs of this Mixture Graph store normals, we can evaluate a shading model at each leaf and propagate the resulting light contributions for diffuse and specular component through the graph to obtain a lookup table for each voxel, across each LOD. Note that this also results in properly pre-filtered lighting contributions, since we do not average normals (as is traditionally done) but light contributions. We would also like to note that estimating smooth normals from voxel-based segmentations is challenging; even more so since data sets from neurosciences typically have strongly anisotropic voxels (up to 1:1:7 for the Neocortex). Fig.~\ref{fig:shading} shows the two lighting modes currently supported by our renderer: on-the-fly computed opacity gradients using discrete differences paired with local ambient occlusion~\cite{Hernell:10} (3-rays, left) and quantized, pre-computed normals stored in a second Mixture Graph (right).

\section{Results}
\hl{Our benchmark configuration comprises a i9-9820X CPU clocked at 3.30GHz and 64GB of RAM, running Windows 10. Although the CPU has 10 cores, our current factorization is mostly sequential and does not utilize the GPU, an NVIDIA RTX Titan with 24GB VRAM.}

\label{sec:results}
\subsection{Factorization}
\begin{table}[ht]
\caption{\label{tab:resultsFactorization}Mixture graph results for the Hippocampus~\cite{Cali16}, Neocortex~\cite{Kasthuri15}, and Polymer~\cite{Weissenboeck:14} data sets using the max-reduction criterion. \hl{Voxel numbers reported include voxels in higher mip-levels.}}
\centering
\footnotesize
\vspace{-1mm}
\begin{tabular}{l|c|c|c}
\def\arraystretch{1.1}
\textbf{Input}          & \textbf{Hippocampus}      & \textbf{Neocortex} & \textbf{Polymer}\\
\hline
resolution              & $1178\times1125\times789$ & $2048^2\times300$ & $614\times 961\times
600$\\
\hl{voxels/million} & \hl{1,195} & \hl{1,438} & \hl{404.7}\\
mip levels              & 12                        & 12 & 11 \\
segments                & 353                       & 1,182 & 15,917\\
background              & 52.24\%                   & 20.70\% & 86.34\% \\
\multicolumn{4}{c}{}\\
\textbf{Histo Mipmap}   & \textbf{Hippocampus}      & \textbf{Neocortex}& \textbf{Polymer} \\
\hline
non-zeros/voxel         & 1.006                     & 1.025 & 1.021\\
raw size                & 9.00 GB                   & 11.10 GB & 3.10GB\\
input node pairs        & 4,976,378                 & 23,551,565 & 184,642,703\\
\multicolumn{3}{c}{}\\
\textbf{Mixture Graph}  & \textbf{Hippocampus}      & \textbf{Neocortex} & \textbf{Polymer} \\
\hline
nodes                   & 1,034,963                 & 5,202,070 & 1,930,257\\
size                    & 1.47 GB                   & 2.14 GB& 0.707 GB \\
bits/voxel              & 11.92                     & 14.52& 17.16\\
quant. bins             & 512                       & 512 & 512\\
max error               & 0.00488                   & 0.00523 & 0.00493\\
topol. depth            & 11                        & 22 & 26\\
encoding time           & 3.51 min                  & 2.05 h & 82.3 h\\
\end{tabular}
\vspace{-3mm}
\end{table}
Table~\ref{tab:resultsFactorization} lists our results for \hl{three} test data sets. The Hippocampus data set \hl{is} provided by Cal\`{i}~\cite{Cali16}. It comprises about 1,045 \hl{million} voxels stemming from electron microscopy imaging, partitioned into 353 segments. \hl{Building} the normalized histogram mipmap of this data set \hl{results} in 1.006 non-zero entries in the mixtures per voxel. \hl{Storing the histogram requires} 9.00~GB. \hl{Subsequent factorization of the} normalized histogram mipmap into the Mixture Graph \hl{results} in slightly more then 1~million nodes and 1.47~GB of raw data. This corresponds to a bitrate of 11.92 bits per input voxel, not considering additional voxels in the mipmap hierarchy.

The second data set is a section of a Neocortex data set~\cite{Kasthuri15}. It was imaged using electron microscopy as well, but contains a much denser segmentation that uses significantly more IDs. Consequently, the total number of node pairs (i.e., the search space for the factorization) in the input is substantially larger. This results in a significantly longer processing time.

The third data set is a micro CT scan of a fiber-reinforced Polymer~\cite{Weissenboeck:14}. Its high segmentation count poses a computational challenge for our current greedy implementation that we plan to address in the near future by relaxing the greed of the algorithm in favour of faster heuristics. \hl{Furthermore, spatially independent factorizations using data windows could lead to increased parallelism.} On the other hand, the data set also shows the greatest potential for the Mixture Graph, since we \hl{are} able to reduce the initial set of over 184.6 million node pairs to just under 2 million, while still maintaining virtually lossless reconstruction.\\

We \hl{quantify} the error introduced by our method as the maximum $\ell_1$ distance between 
\hl{the original $\His$ and a full reconstruction from the Mixture Graph. We obtain the reconstruction by assigning unit vectors $\unit{i}$ to leafs $i$ followed by propagation. Errors are thus relative to a unit of $1$ and dimensionless.
} 
The reason for that choice is that all mixtures are $\ell_1$-normalized. Note that, unlike the error estimate discussed in Section~\ref{sec:error}, the error listed in \hl{Table~{\ref{tab:resultsFactorization}} (max error) is an actual error based on a encoding-decoding round trip. In contrast, the error estimate quantifies the uncertainty} in each ID as required for volumetric queries.\\

In addition to the number of nodes in the Mixture Graph (including sources and sinks), we also list the topological depth of the graph, since it is equivalent to the number of synchronization steps during parallel reconstruction. \hl{We provide} timings for \hl{all} data sets \hl{including} generation of the histogram mipmap and the factorization into the Mixture Graph.\\

For all data sets, we observe a significant reduction in the computational workload required to propagate transfer function updates through the entire mipmap. This is reflected by the ratio between voxels in the input mipmap (Hippocampus: 1,195M, Neocortex: 1,438M, Polymer: 404.7M) and the generated nodes (Hippocampus: 1M, Neocortex: 5.2M, Polymer: 1.9M). While the number of generated nodes depends on the number of possible pairs to choose from (Hippocampus: ~5M, Neocortex: 23.6M, Polymer: 180.6M), it also depends on the number and location of segments. It is therefore no surprise that the Polymer data with its many fibrous segments shows the best reduction since mixtures involving many segments are quite rare. In any case, the reduction is substantial ($>~4:1$ and almost $100:1$ for the Polymer). Moreover, without the Mixture Graph, hundreds of millions or even billions of voxels \hl{have} to be revisited on every transfer function change,\\

\noindent\textbf{Max-occurrence criterion.} Both the max-occurrence criterion (Eq.~\ref{eq:greedy1}) and the max-reduction criterion (Eq.~\eqref{eq:greedy2}) \hl{result} in a similar result regarding output size and max error. However, the max-reduction criterion \hl{is} significantly faster in \hl{all three} cases (around $2.5\times$ for the Hippocampus and around $2.7\times$ for Neocortex and Polymer), which we attribute to the fact that it prunes the search space quicker. However, to our surprise, the max-reduction criterion also \hl{results} in a slightly better error, albeit negligibly so. The max-reduction criterion also \hl{generates} a negligibly ($\leq$ 1 permille) smaller number of nodes.\\

\noindent\textbf{Quantization parameter.} We also conduct experiments to assess the impact of the scalar quantization stage on the overall performance of our method. Our findings are summarized in Table~\ref{tab:coding}. \hl{The error reported there is obtained by taking the maximum of the each quantization bin's individual rmse. Since quantization bins store interpolation weights, the error is in $\left[0,1\right]$.} 
Lower bitrates in the scalar quantization stage lead to higher ambiguities between nodes, and, thus, to substantially fewer nodes. To see this, consider that two interpolations $\lambda(i,j,w)$, $\lambda(i,j,w^\prime)$ are merged if the quantization bin of $w$ is identical to that of $w^\prime$. Fewer nodes translate to faster encoding and reconstruction times. The cost to pay is that the reconstruction error roughly doubles for each reduction by one bit. 
\hl{The reason is} that each such reduction reduces the precision of the quantized interpolation weight by half. Seemingly surprising, the output size stays more or less constant. This is a consequence of the fact that the bulk of the output size is allotted to voxel data at the finest level. Interestingly, the encoding time decreases a little faster than the number of nodes. We attribute this to our observation that, irrespective of the bitrate, similar choices are made in the early stages of our greedy algorithm. That means that for any bitrate, the search space is reduced by roughly the same absolute amount early on, resulting in fewer factorization steps at lower bitrates.\\

\begin{table}
\captionsetup{skip=2pt}
\caption{\label{tab:coding}Impact of the bitrate of the scalar quantizer.}
\centering
\footnotesize
\begin{tabular}{c|r|r|r|r}
 ~ & \multicolumn{4}{c}{\textbf{Hippocampus}}\\
rate/bits & nodes & size & time & error\\
\hline
4 & 460,210   & 1.447 GB & 2.21 min & 0.17475\\
5 & 655,225   & 1.467 GB & 2.51 min & 0.09384\\
6 & 809,324   & 1.469 GB & 2.75 min & 0.04213\\
7 & 898,406   & 1.469 GB & 2.92 min & 0.02525\\
8 & 968,962   & 1.469 GB & 3.21 min & 0.01228\\
9 & 1,029,188 & 1.470 GB & 3.51 min & 0.00488\\
\multicolumn{5}{c}{}\\[-2ex]
~ &\multicolumn{4}{c}{\textbf{Neocortex}} \\
rate/bits & nodes & size & time & error\\
\hline
4 & 3,286,086 & 2.102 GB & 90.25 min & 0.23748\\
5 & 3,980,659 & 2.107 GB & 95.36 min & 0.10286\\
6 & 4,441,354 & 2.132 GB & 102.70 min & 0.05272\\
7 & 4,758,860 & 2.134 GB & 106.37 min & 0.02639\\
8 & 5,038,103 & 2.136 GB & 116.26 min & 0.01288\\
9 & 5,300,258 & 2.139 GB & 127.20 min & 0.00523\\
\multicolumn{5}{c}{}\\[-2ex]
~ &\multicolumn{4}{c}{\textbf{Polymer}} \\
rate/bits & nodes & size & time & error\\
\hline
4 & 1,071,148 & 713.6 MB & 71.6 h & 0.18248\\
5 & 1,315,833 & 718.3 MB & 73.2 h & 0.07603\\
6 & 1,519,865 & 721.5 MB & 76.6 h & 0.04066\\
7 & 1,684,252 & 722.6 MB & 79.5 h & 0.01932\\
8 & 1,812,057 & 723.1 MB & 80.7 h & 0.01001\\
9 & 1,930,257 & 724.0 MB & 82.3 h & 0.00493\\
\end{tabular}
\vspace{-4mm}
\end{table}

\subsection{Reconstruction}
\hl{The CPU in our benchmark configuration} supports bit manipulation instructions for \hl{counting leading and trailing zeros} that result in an easier implementation of the footprint gathering algorithm.\\

\begin{figure}[!htb]
\centering
    \includegraphics[width=0.32\columnwidth]{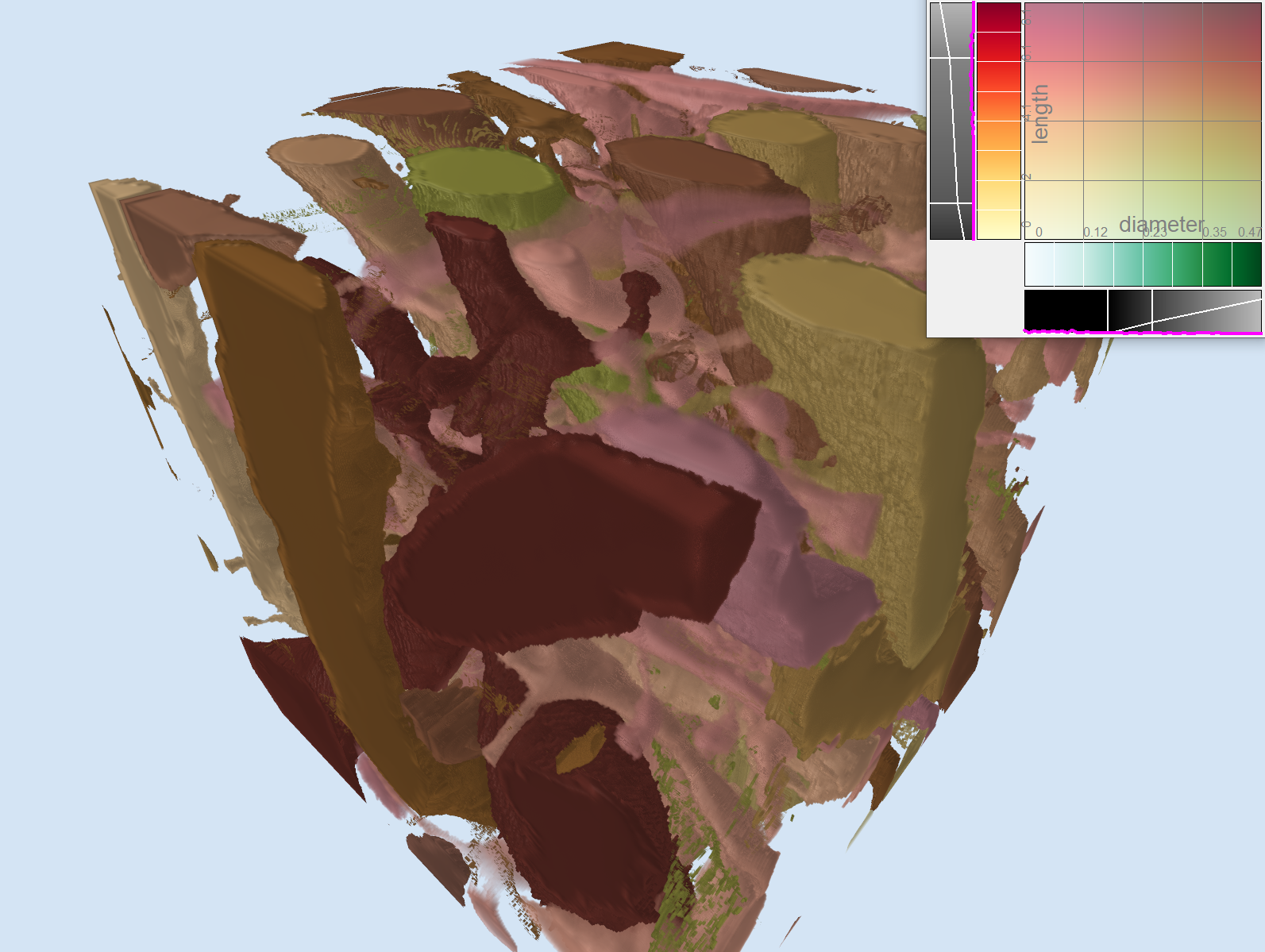}
    \includegraphics[width=0.32\columnwidth]{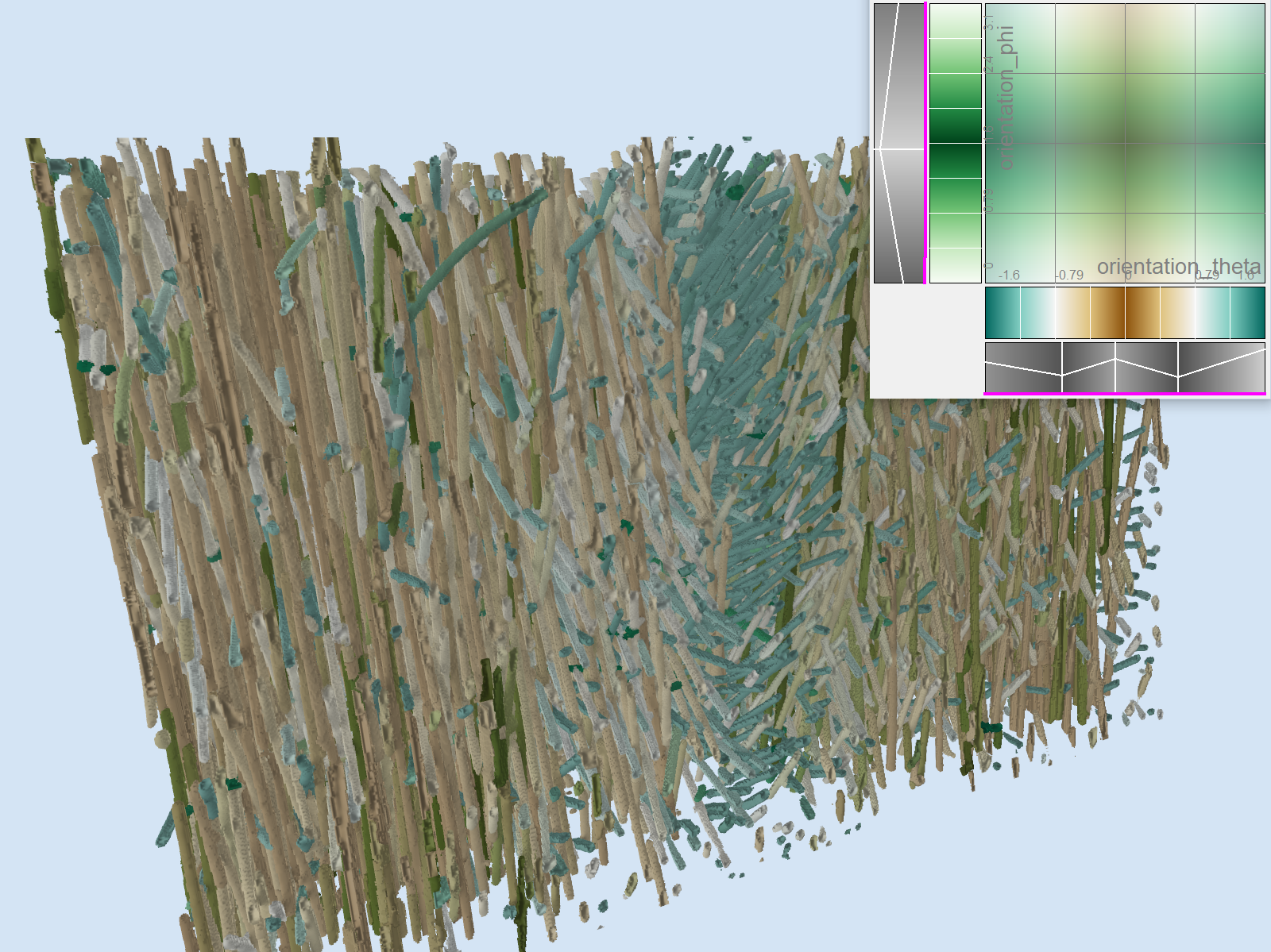}
	\includegraphics[width=0.32\columnwidth]{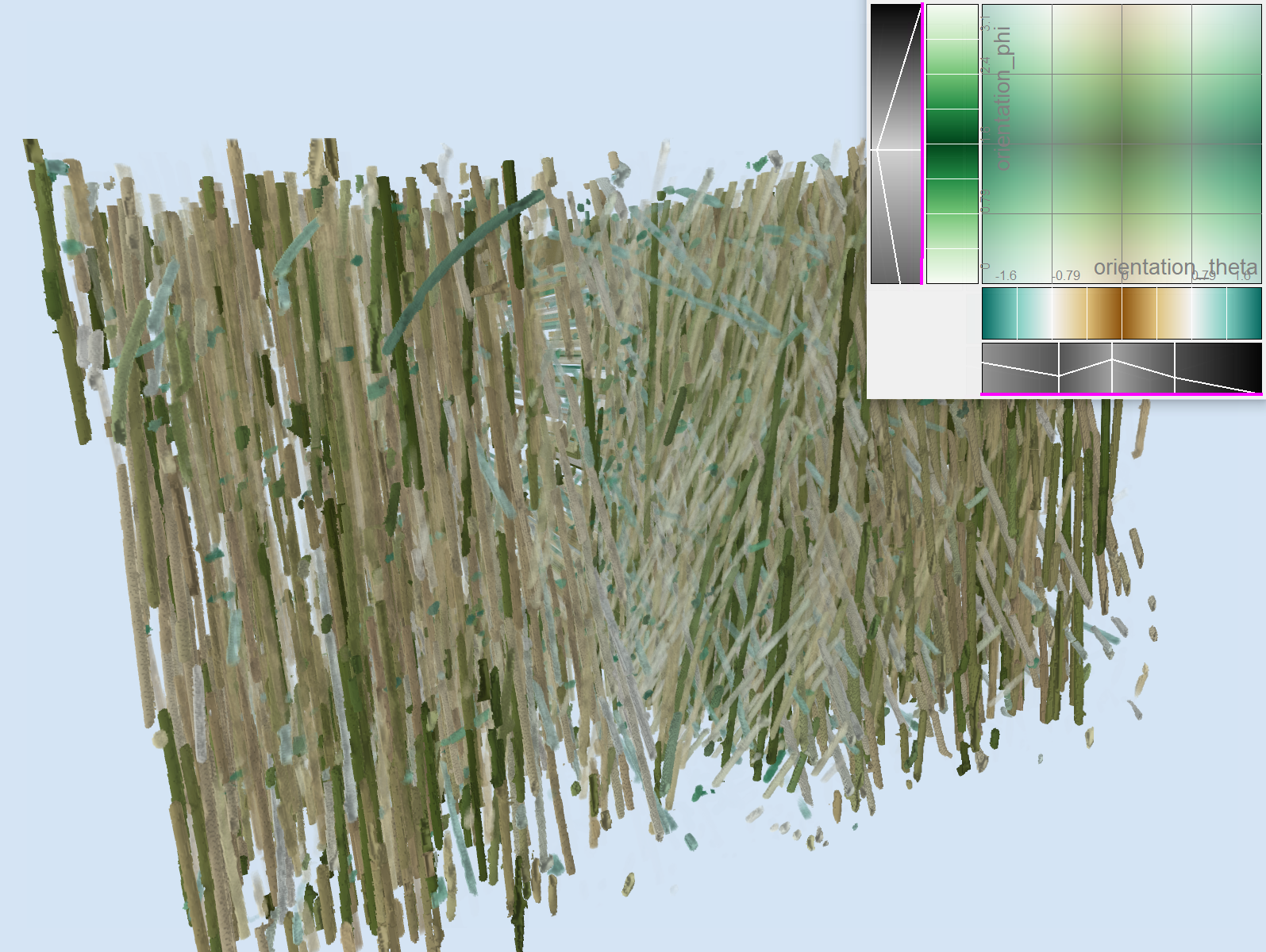}
	\vspace{-2mm}
	\caption{\label{fig:transfer-function}The Mixture Graph allows for real-time transfer function editing across all scales of the mipmap, and can be used for multidimensional color mapping of different segment attributes. These schemes can be used for various real time visual analysis applications in neuroscience~(left), or material science~(center and right).
	\vspace{-2mm}
	}
\end{figure}
\noindent\textbf{Color lookup table.} \hl{We evaluate} the time it takes to assemble a color lookup table on both CPU and GPU. \hl{To do so, we assign} colors to the source leaf nodes in the DAG and traverse all remaining nodes in parallel. For each node, we fetch data from its children and perform one interpolation. 
\hl{We measure the execution time of 1,000 repetitions and report the average time of one repetition} and GPU timings include the data upload of one float4 color per source.
 On the CPU, we used OpenMP to parallelize the code, whereas on the GPU, a compute shader is used. This code is executed whenever the user changes the color transfer function such as shown in Fig.~\ref{fig:transfer-function}.
For the Hippocampus data set (353 sources, topological depth 11) our code \hl{takes}
22.7ms on the CPU and 3.5ms on the GPU for computing color information for the more than 1M nodes. For the Neocortex (1,182 sources, topological depth 22) our code \hl{takes} 129.7ms (CPU) and 11.3ms (GPU). Finally, for the polymer (15,917 sources, topological depth 26) our code \hl{takes} 57.0ms (CPU) and 12.1ms (GPU). These timings indicate that the limiting factor on the CPU is the number of nodes whereas on the GPU the need to synchronize affects computation times. 
\hl{Enabled by this performance, we implement several interactive multidimensional transfer function editing schemes that can be easily integrated into a wide range of visual exploration workflows. For example, neuroscientists can classify and select neural structures according to geometric features such as shape, volume, length and diameter~(Fig.~{\ref{fig:transfer-function}} left), and material scientists can cluster fibers into groups based on length or orientation~(Fig.~{\ref{fig:transfer-function}} middle and right).}\\

\begin{figure}[b]
\centering\vspace{-4mm}
	\includegraphics[width=0.8\columnwidth]{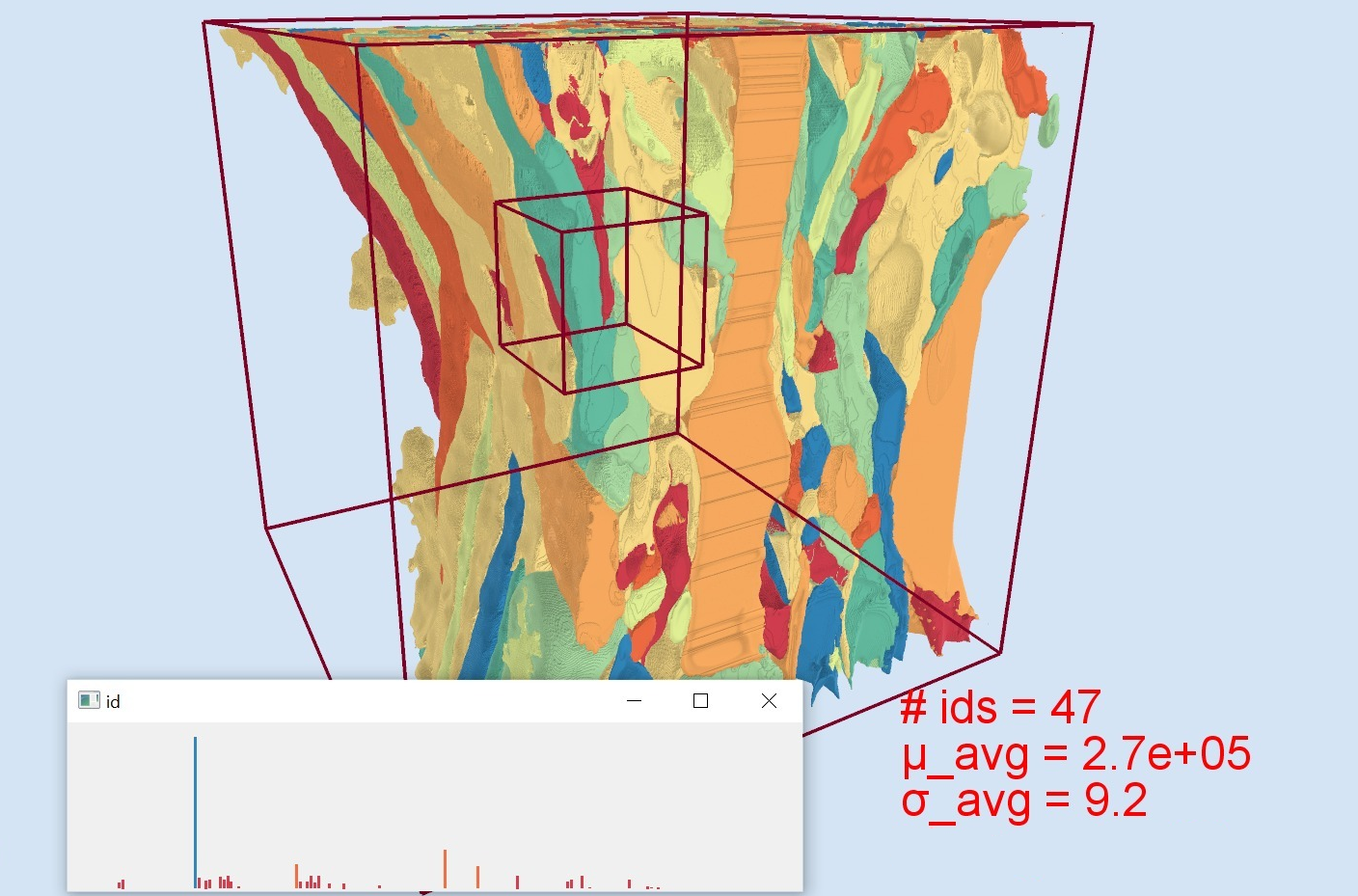}
	\vspace{-2mm}
	\caption{\label{fig:FAscreen}The Mixture Graph allows for real-time computation of approximate histograms over axis-aligned bounding boxes. The above small box contains segments with an average volume of $270K\pm9.2$ voxels per segment. 
	\vspace{-1mm}
	}
\end{figure}

\noindent\textbf{Mixture lookup table.} \hl{We also bench} the time it takes to assemble a mixture lookup table on the CPU. \hl{We use} this lookup table for range queries (both na\"{i}ve and our footprint assembly algorithm), such as depicted in Fig.\ref{fig:FAscreen}. We start by assigning a mixture $\unit{i}$ with variance 0 to each leaf node $i$. In the same fashion as for the color lookup, we propagate this information through the DAG, including the error described in Eq.~\eqref{eq:error}. Each node thus stores two sparse vectors over $\R^\infty$, the expected value $\mu$ and its variance $\delta\mu$. For the Hippocampus, computing this lookup takes 68ms, 389ms for the Neocortex, and 171ms for the Polymer. 
\begin{figure*}[t]
\centering
	\includegraphics[width=0.185\textwidth]{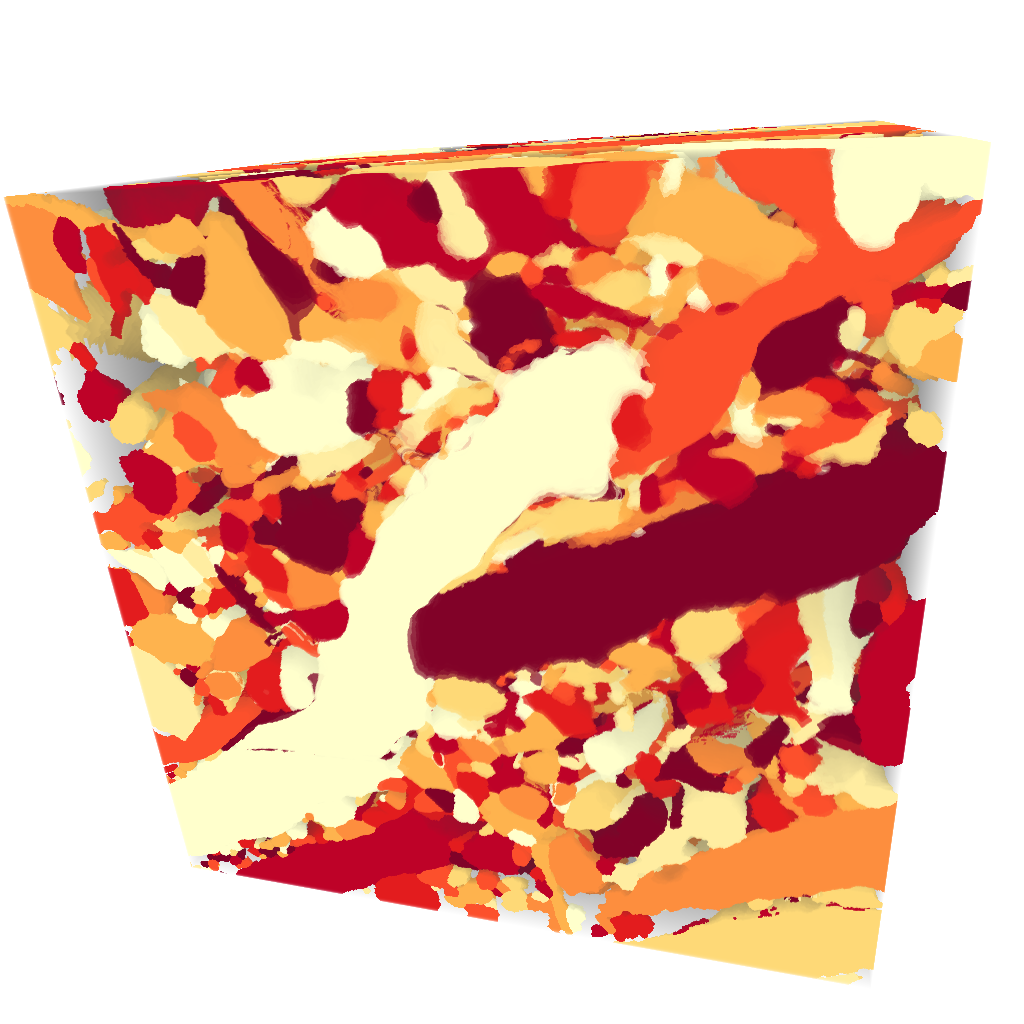}
	\includegraphics[width=0.185\textwidth]{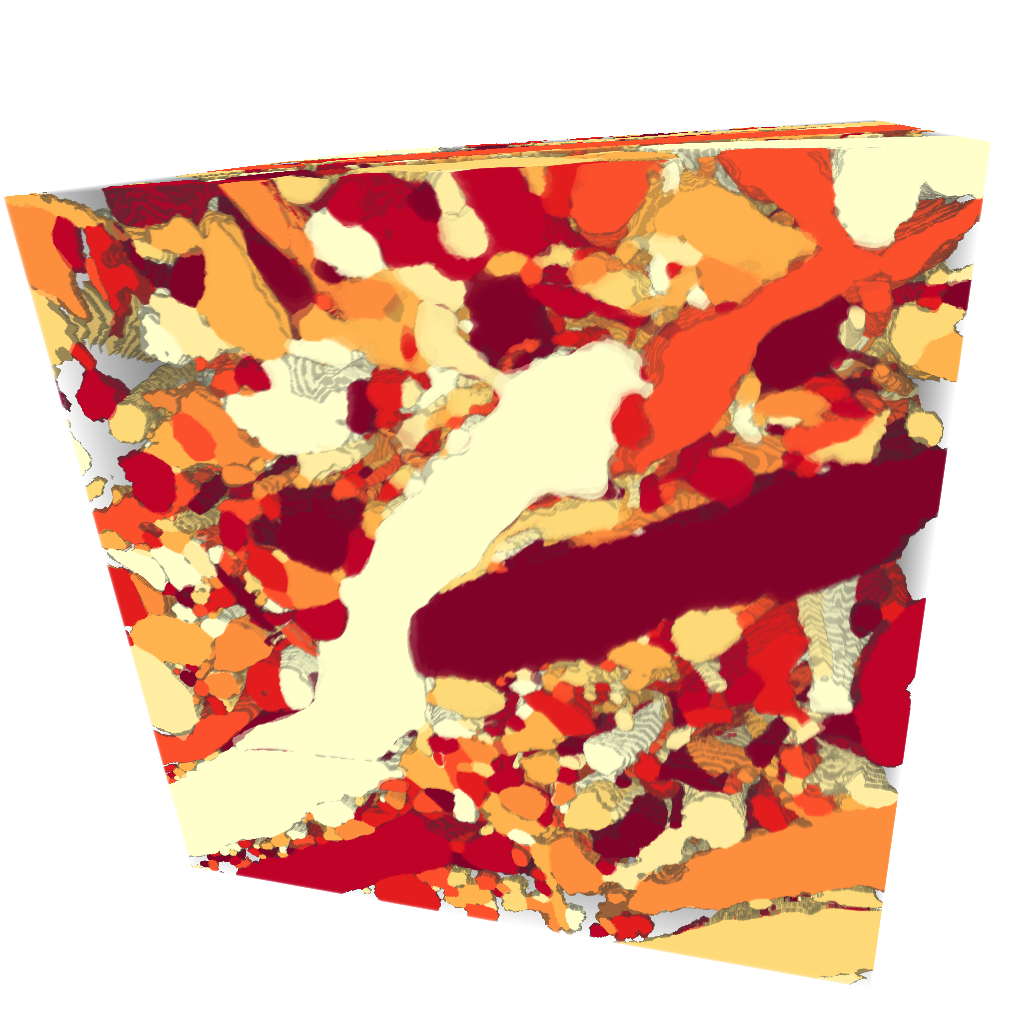}
	\includegraphics[width=0.185\textwidth]{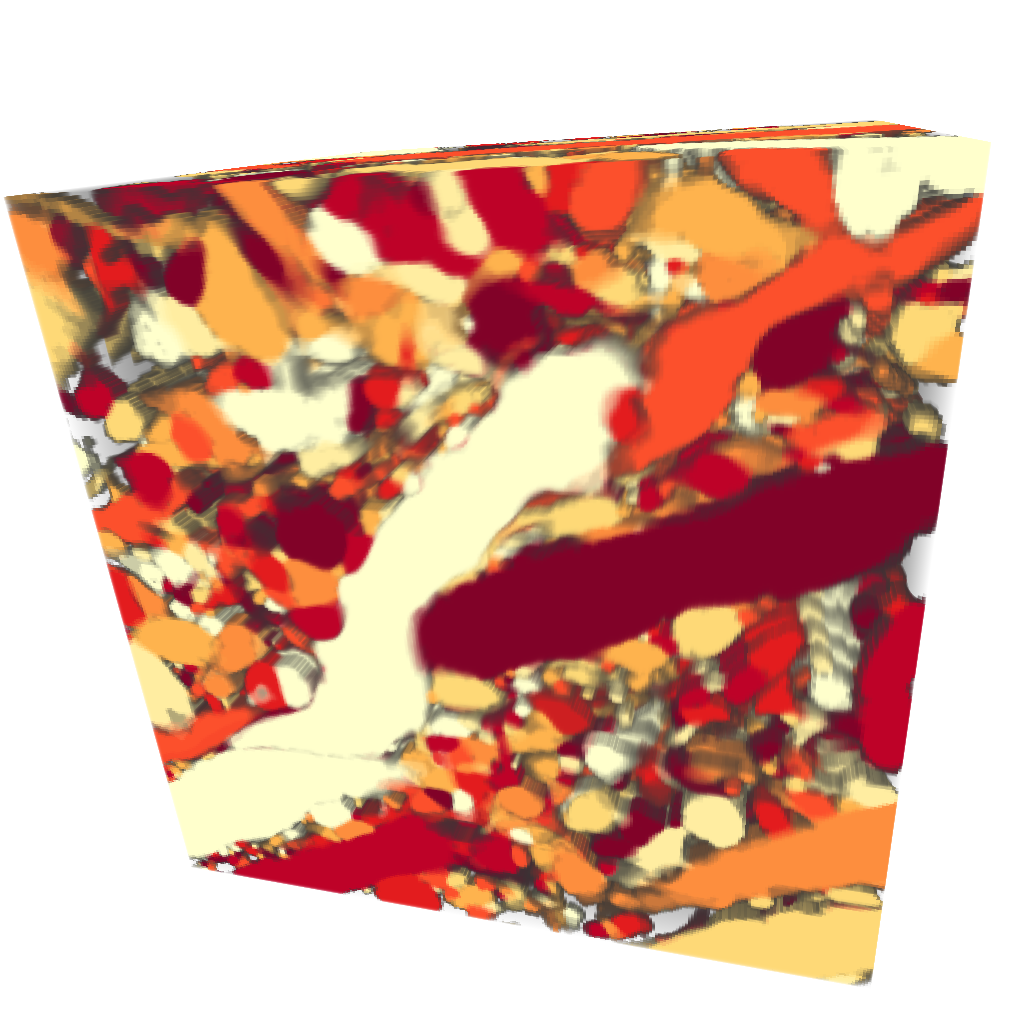}
	\includegraphics[width=0.185\textwidth]{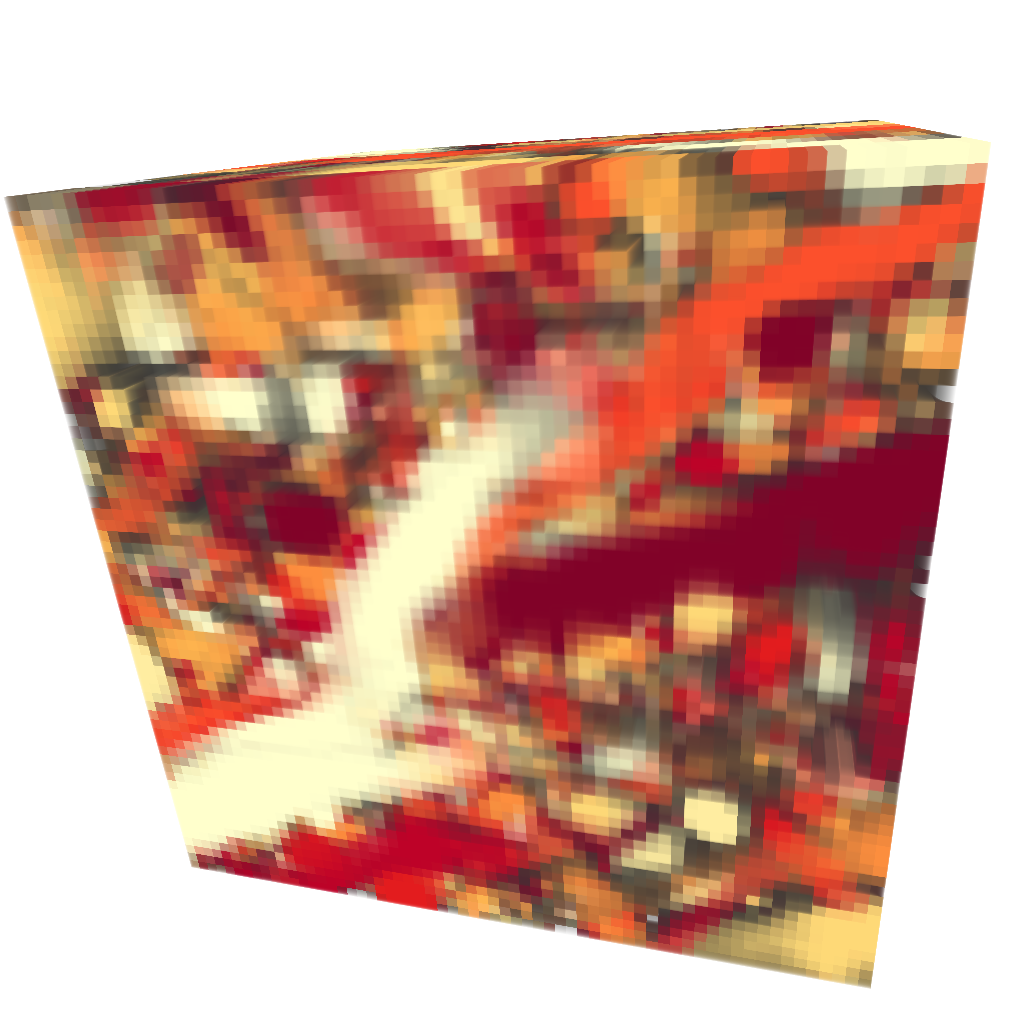}
	\includegraphics[width=0.185\textwidth]{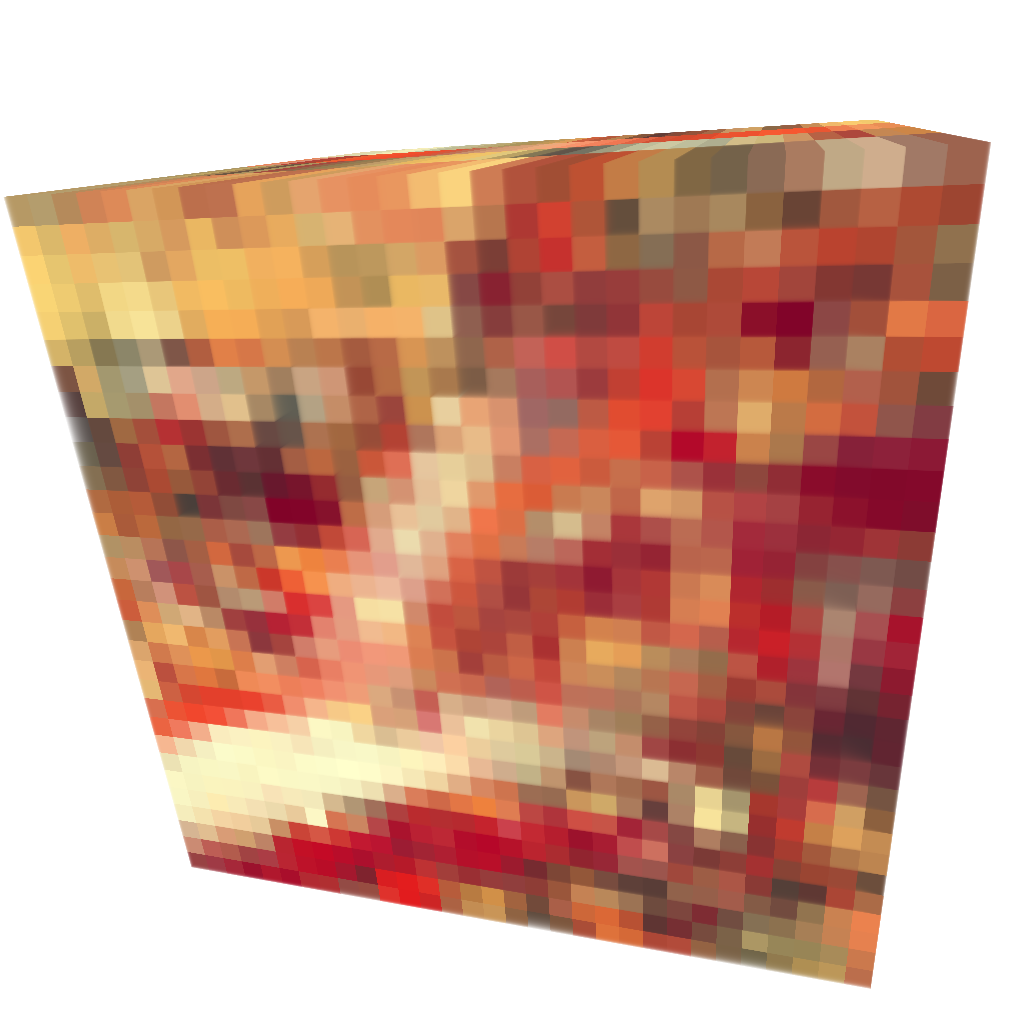}\\
	\includegraphics[width=0.185\textwidth]{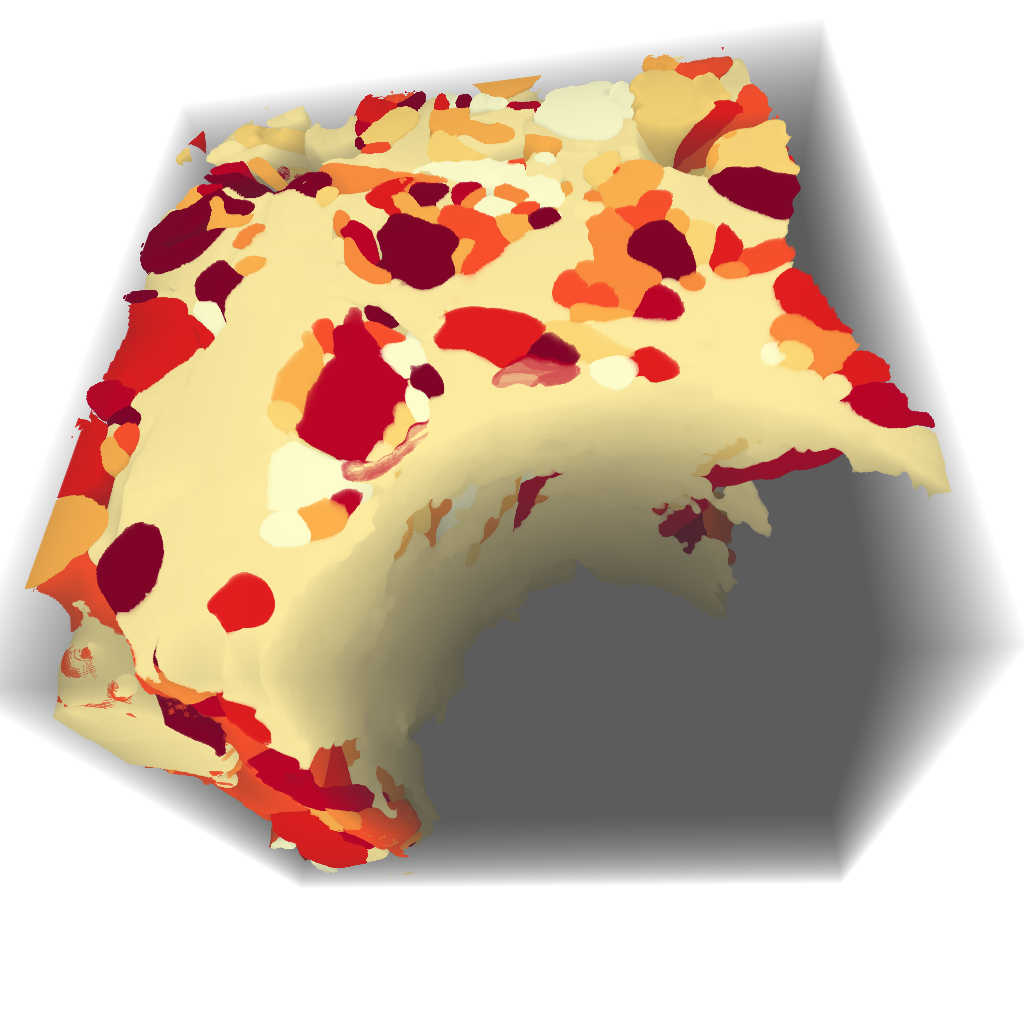}
	\includegraphics[width=0.185\textwidth]{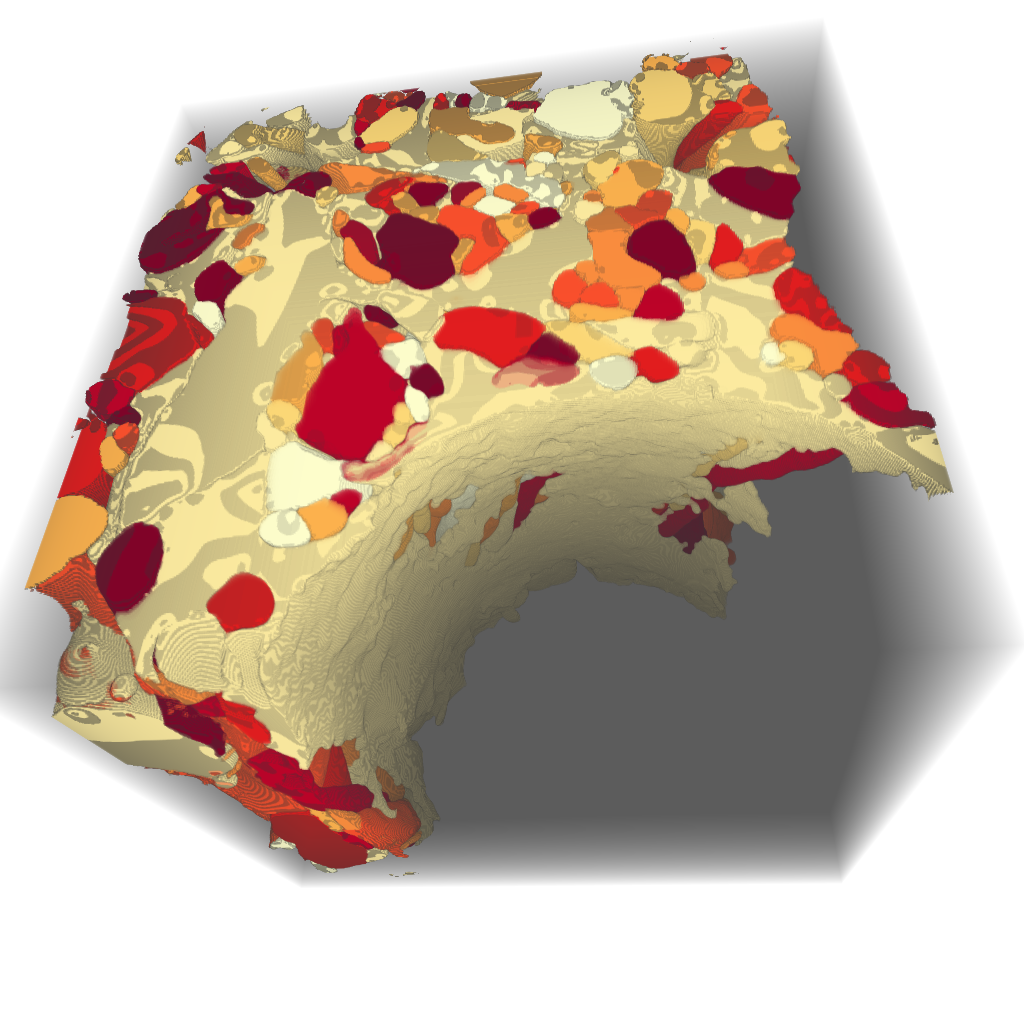}
	\includegraphics[width=0.185\textwidth]{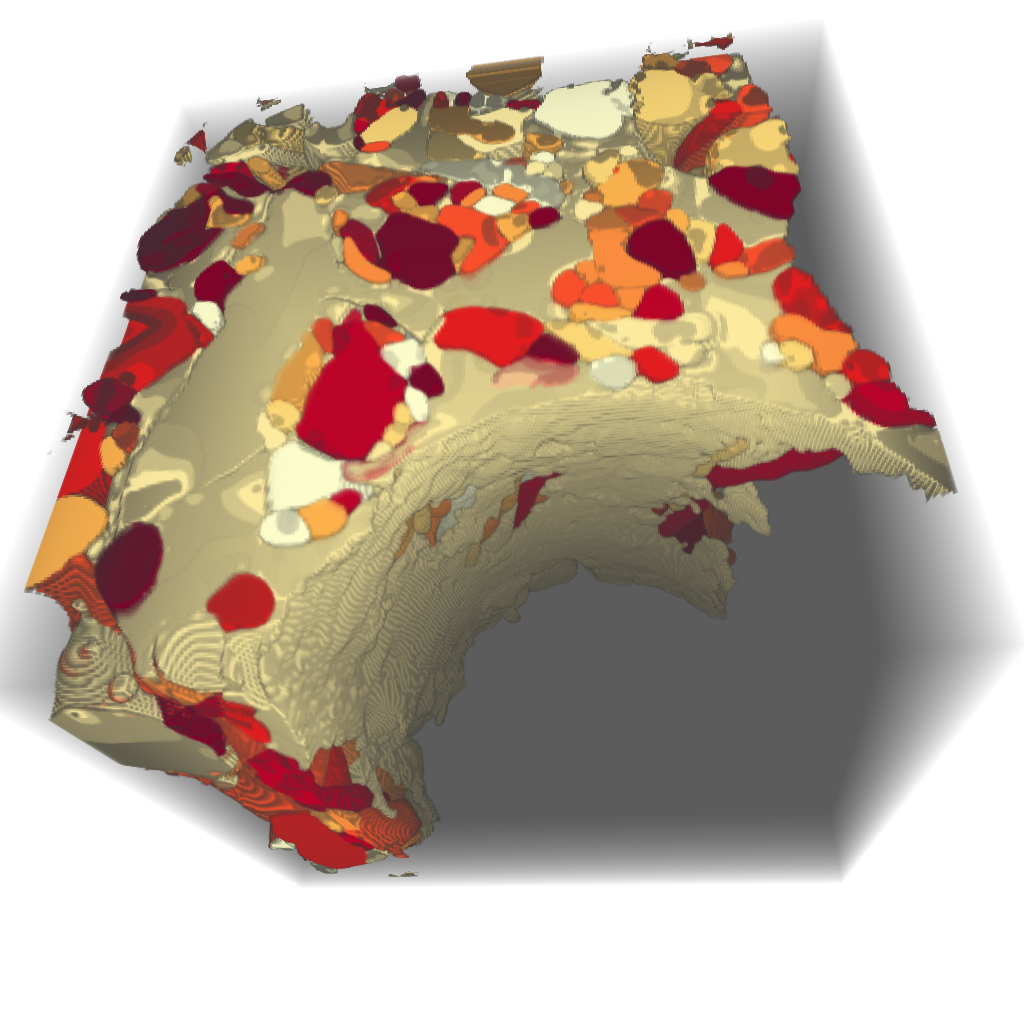}
	\includegraphics[width=0.185\textwidth]{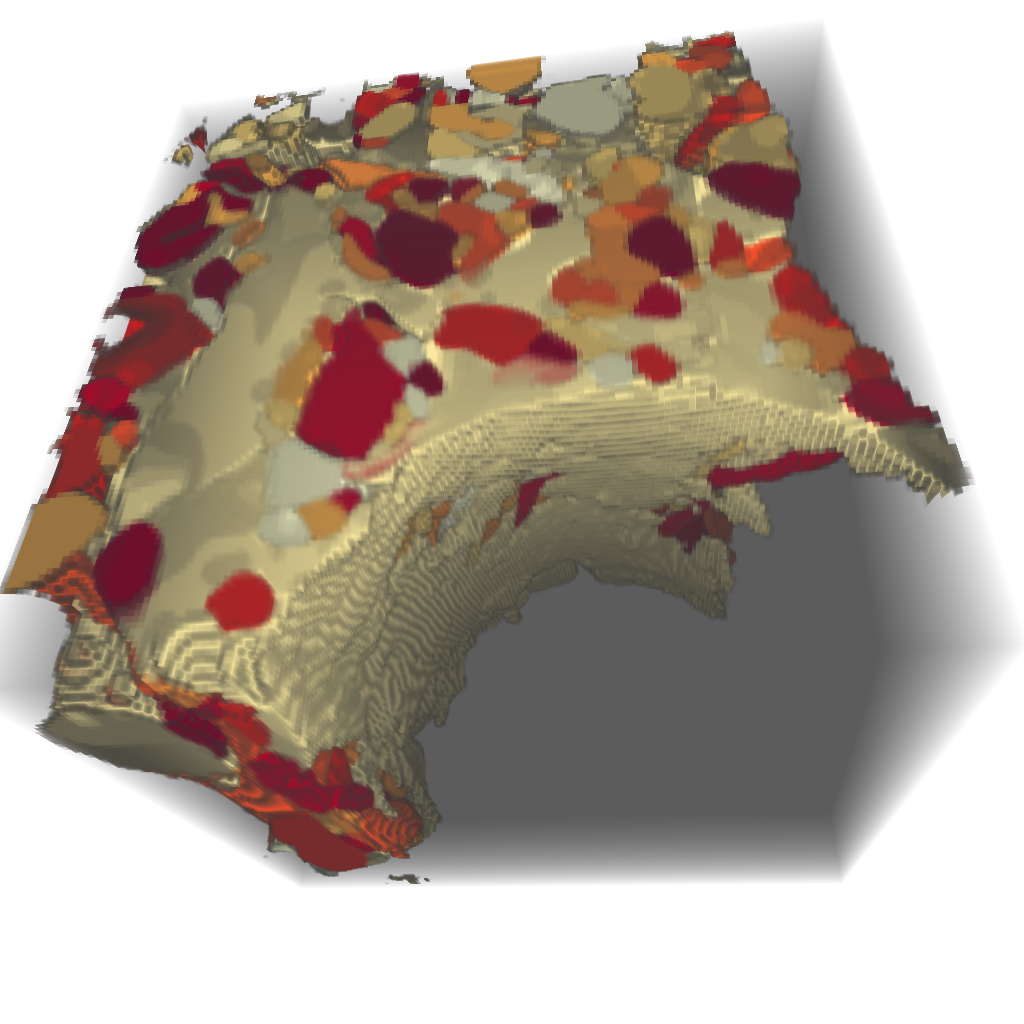}
	\includegraphics[width=0.185\textwidth]{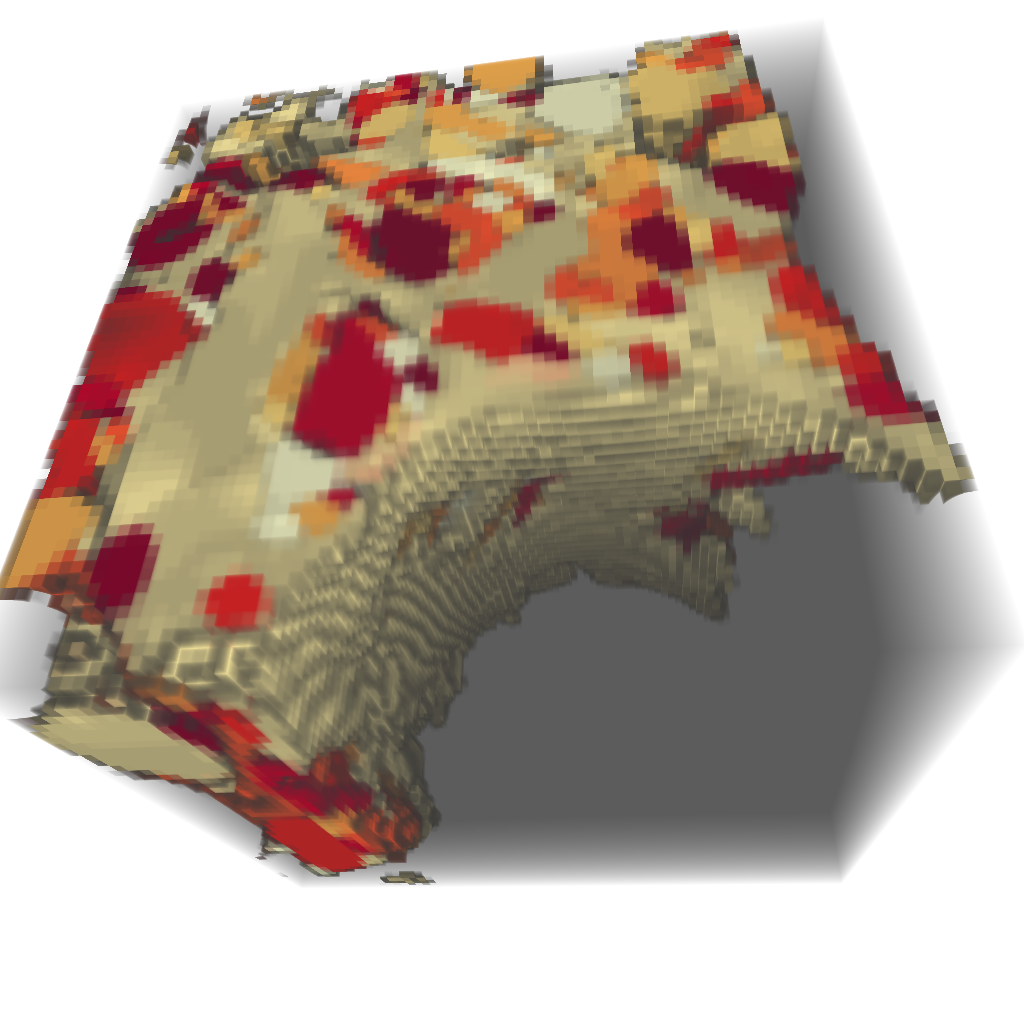}
	\vspace{-4mm}
	\caption{\label{fig:rendering}\textbf{Top, left to right:} Direct volume rendering of levels 0,2,4,5, and 6 of the 1.2G voxel Neocortex data set. The data sets strong anisotropy can be seen in the middle images. 
	\textbf{Bottom, left to right:} Direct volume rendering of levels $0,\ldots,4$ of the 1.0G voxel Hippocampus data set.
	For both rows, empty regions were set to a semi-transparent gray, and shading was disabled for a cleaner illustration.
	These images were rendered directly from the binary Mixture Graph representation at interactive rates on the GPU.
	\vspace{-1mm}
	}
\end{figure*}
Building this lookup table does not only depend on the number of nodes and the topological depth of the DAG, but also on the initial number of segments as well as the number of non-zero entries in the sparse vectors. This operation will typically be performed only once when loading the data set. Due to the pruning performed during the factorization, the lookup tables easily fit into main memory, with the Neocortex lookup at around 251MB being the largest of the three.\\

\noindent\textbf{Range queries.} We then \hl{use} this lookup table to measure randomly-aligned range queries covering between $16^3$ to $256^3$ voxels. For each of the \hl{three} data sets, we compare a parallel na\"{i}ve approach that adds the contributions of each voxel at the finest level with our footprint assembly (FA) method that exploits the data hierarchy. Our findings are summarized in Table~\ref{tab:query}. As can be seen, FA significantly reduces the \hl{number} of samples necessary to count the segment volumes in a given range. At ranges larger than $8^3$, we amortize the setup overhead of FA over the na\"{i}ve method. Beyond this point, we observe substantial speed-ups that are enabled by our hierarchical data structure.

\begin{table}[htb!]
\captionsetup{skip=0pt}
\caption{\label{tab:query} \hl{Performance of the footprint assembly (FA) algorithm: \\ \textbf{Hippocampus} (top), \textbf{Neocortex} (center), and \textbf{Polymer} (bottom).
Left to right: size of the query region, fetches made by the FA algorithm, percentage of fetches saved, time in ms for FA and 10-core parallel {na\"{i}ve} approach, and speedup (bold times in seconds). }}
\centering
\begin{tabular}{c|r|r|r|r|r}
\multicolumn{3}{l}{} & \multicolumn{2}{|c|}{time/ms}\\
size & fetches & saved & footprint & na\"{i}ve & speed-up\\
\hline
 $16^3$ & 768.6& 81.24\% & 0.1136 & 0.4058 & 3.57$\times$ \\
 $32^3$ & 3.92K & 88.04\% & 0.3597 & 1.8547 & 5.16$\times$ \\
 $64^3$ & 17.45K & 93.34\% & 1.3577 & 12.8862 & 9.49$\times$ \\
 $128^3$ & 73.9K & 96.48\% & 6.0933 & 110.5094 & 18.14$\times$ \\
 $256^3$ & 264.7K& 98.42\% & 25.2157 & \textbf{1.2s} &  57.60$\times$ \\
\hline \hline
 $16^3$ & 961.40 & 76.53\% & 0.1463 & 0.4015 & 2.74$\times$ \\
 $32^3$ & 3.9K & 87.96\% & 0.4091 & 1.8542 & 4.53$\times$ \\
 $64^3$ & 12.7K & 95.15\% & 1.2531 & 14.5791 & 11.63$\times$ \\
 $128^3$ & 63.1K & 96.99\% & 7.7496 & 160.4324 & 20.70$\times$ \\
 $256^3$ & 246.8K & 98.53\% & 89.5613 & \textbf{2.6s} & 28.98$\times$ \\
\hline \hline
 $16^3$ & 925.70 & 77.40\% & 0.1344 & 0.4359 & 3.24$\times$ \\
 $32^3$ & 3.9K & 88.09\% & 0.3912 & 2.4141 & 6.17$\times$ \\
 $64^3$ & 17.9K & 93.15\% & 2.1309 & 29.5320 & 13.86$\times$ \\
 $128^3$ & 69.4K & 96.69\% & 15.8873 & 652.8276 & 41.09$\times$ \\
 $256^3$ & 289.6K & 98.27\% & 393.9230 & \textbf{70.5s} & 178.95$\times$ \\
\end{tabular}
\vspace{-2mm}
\end{table}
It is worth noting that the number of fetches made by the FA method depends solely on the domain resolution, as well as on the footprint size and alignment. These numbers are therefore similar for the \hl{three} data sets. In contrast, the time is also affected by the number of segmentation IDs and their spatial distribution. For large, contiguously labeled regions, the resulting mixture remains very sparse throughout its computation, whereas for regions with less coherence, many more terms have to be accumulated.

\subsection{Rendering}

\hl{A drawback of our method to derive pre-computed normals are artifacts in the form of \textit{caps} around the medial axis where the segments leave the domain.} A second limitation is that we are restricted to light and camera at infinity since we compute lighting contributions without knowledge of position. Note that only the second is a limitation of the Mixture Graph since our method is orthogonal to the normal estimation. Fig.~\ref{fig:rendering} depicts several levels of the Hippocampus and Neocortex data sets, all rendered at the same resolution, to illustrate the downsampling capabilities of our data structure. In the accompanying video, we demonstrate real-time performance using a direct volume raymarcher, sampling at 0.5 voxels and rendering to a $1600\times 1200$ viewport.

\subsection{Conclusion \& Future Work}

In this paper, we have presented the Mixture Graph. This data structure allows us to factorize and compress normalized histogram mipmaps, essentially resulting in a paletted mipmap that provides fast updates of transfer functions for rendering. In our experiments, we \hl{observe} a decrease from the order of gigavoxels in the mipmap down to the order of mega-nodes in the Mixture Graph. Both quantities reflect the amount of computational work required to re-compute the mipmap. We \hl{demonstrate} the usefulness of our method for segmented volumes, whose integer nature otherwise prohibit direct filtering, interpolation, and lossy compression. Additionally, we \hl{evaluate and present} various trade-offs between encoding speed and reconstruction fidelity to guide future users of our method.\\

Our current implementation considers a single volume and builds the Mixture Graph using serial code. In the future, we would like to explore bricked volumes, since they offer potential for parallelization and out-of-core processing. However, building the Mixture Graph of a bricked volume in parallel is not straightforward. The reason is that each factorization step alters the remaining search space and, consequently, the order of execution may matter significantly. Another direction for future research is to consider larger mixtures (e.g., barycentric interpolation etc.) as the building blocks of the graph. While each mixture would require more storage, we expect that fewer nodes will be generated. Also, the longest path in the graph is expected to become shorter, which would require fewer synchronization operations during parallel transfer function updates. \hl{Our current empty space skipping method is efficient but we also plan to investigate how more sophisticated schemes~{\cite{Hadwiger2018}} can be combined with out data structure in the future.}
Finally, we \hl{demonstrate} that the choice of the greedy scoring function has significant impact on the performance of the construction of the mixture graph. In this context, more research is needed to determine better scoring functions.

\acknowledgments{
All authors are funded by the College of Science and Engineering (CSE) at Hamad Bin Khalifa University (HBKU).
The data sets used in this work were generated by Cal\`{i} et al.~\cite{Cali16} (Hippocampus) and Kasthuri et al.~\cite{Kasthuri15} (Neocortex). The fiber-reinforced polymer data set \hl{is} provided by Christoph Heinzl~\cite{Weissenboeck:14}}. 

\note

\clearpage
\bibliographystyle{abbrv}
\bibliography{LDAG}

\end{document}